\documentclass[pra,twocolumn,showkeys,showpacs,superscriptaddress,floatfix,longbibliography]{revtex4-1}

\usepackage[T1]{fontenc}
\usepackage[latin9]{inputenc}
\usepackage{epsfig,psfrag}
\usepackage{latexsym}
\usepackage{graphicx}
\usepackage{dcolumn}
\usepackage{amssymb}
\usepackage{amsmath}
\usepackage{bm}
\usepackage{mathrsfs}
\usepackage{bigints}
\usepackage{upgreek}
\usepackage{color}
\usepackage{longtable}
\usepackage{xspace}
\usepackage[squaren]{SIunits}
\usepackage{epstopdf}
\usepackage[normalem]{ulem}

\definecolor{mygreen}{RGB}{20,148,20}

\usepackage{filemod}

\newcommand{\ie}{i.e.\@\xspace}

\newcommand{\eq}[1]{Eq.~\eqref{eq:#1}}
\newcommand{\eqs}[2]{Eqs.~\eqref{eq:#1} and~\eqref{eq:#2}}
\newcommand{\eqr}[2]{Eqs.~\eqref{eq:#1} to~\eqref{eq:#2}}

\newcommand{\fig}[1]{Fig.~\ref{fig:#1}}

\renewcommand{\bm}[1]{\boldsymbol{\mathbf{#1}}}
\newcommand{\msisa}{\Sigma_\text{ISA}}
\newcommand{\mscor}{\Sigma_\text{cor}}
\newcommand{\msrec}{\Sigma_\text{rec}}
\newcommand{\mvisa}{\tensor{\boldsymbol{\Sigma}}_\text{ISA}}
\newcommand{\mvcor}{\tensor{\boldsymbol{\Sigma}}_\text{cor}}
\newcommand{\mvrec}{\tensor{\boldsymbol{\Sigma}}_\text{rec}}
\newcommand{\mtisa}{\Sigma_{t,\text{ISA}}}
\newcommand{\mtcor}{\Sigma_{t,\text{cor}}}
\newcommand{\mtrec}{\Sigma_{t,\text{rec}}}
\newcommand{\mstot}{\Sigma}
\newcommand{\mvtot}{\tensor{\boldsymbol{\Sigma}}}
\newcommand{\mttot}{\Sigma_t}

\newcommand{\comm}[1]{}
\newcommand{\reply}[1]{}

\begin{document}

\title{Coherent light propagation through cold atomic clouds beyond the independent scattering approximation}

\author{C. C. Kwong}
\affiliation{School of Physical and Mathematical Sciences, Nanyang Technological University, 637371 Singapore, Singapore.}
\affiliation{MajuLab, CNRS-UCA-SU-NUS-NTU International Joint Research Unit, Singapore}
\author{D. Wilkowski}
\affiliation{School of Physical and Mathematical Sciences, Nanyang Technological University, 637371 Singapore, Singapore.}
\affiliation{MajuLab, CNRS-UCA-SU-NUS-NTU International Joint Research Unit, Singapore}
\affiliation{Centre for Quantum Technologies, National University of Singapore, 117543 Singapore, Singapore}
\author{D. Delande}
\affiliation{Laboratoire Kastler Brossel, Sorbonne Universit\'{e}, CNRS,
	ENS-Universit\'{e} PSL, Coll\`{e}ge de France, 4 Place Jussieu, 75005 Paris, France}
\author{R. Pierrat}
\affiliation{Institut Langevin, ESPCI Paris, CNRS, PSL University, 1 rue Jussieu, 75005 Paris, France}

%\date{\filemodprint{\jobname}~~File: \jobname}

\begin{abstract}
    We calculate the relative permittivity of a cold atomic gas under weak probe illumination, up to second order in the density. Within the framework of a diagrammatic representation method, we identify all the second order diagrams that enter into the description of the relative permittivity for coherent light transmission.  These diagrams originate from pairwise position correlation and recurrent scattering. Using coupled dipole equations, we numerically simulate the coherent transmission with scalar and vector waves, and find good agreement with the perturbative calculations. We applied this perturbative expansion approach to a classical gas at rest, but the method is extendable to thermal gas with finite atomic motion and to quantum gases where non-trivial pair correlations can be naturally included.
\end{abstract}

\pacs{}

\keywords{}

\maketitle

\section{Introduction}
% ====================

Cold atomic systems have been used to study various phenomena related to the scattering and coherent transport of light in disordered diluted media such as radiation trapping~\cite{fioretti1998,labeyrie2003}, coherent backscattering~\cite{labeyrie1999,bidel2002}, random lasing ~\cite{baudouin2013}, and (super)flash effect~\cite{chalony2011,kwong2014cooperative,kwong2015cooperative,bettles2018faster}. A lot of experimental efforts have also been initiated to understand wave transport in dense atomic media, when the light scattering mean free path becomes comparable or even smaller than the wavelength of light. In this case, the independent scattering approximation (ISA) is expected to break down and complex collective or cooperative mechanisms emerge.
As such, signatures of light localization~\cite{balik2013, sokolov2015}, collective emission of light like superradiance and subradiance~\cite{bienaime2012,araujo2016,roof2016,araujo2017} have been reported.

In parallel to experimental progresses, numerous theoretical models and numerical simulations has been developed to understand the scattering and transmission of light in dense media~\cite{sokolov2009,sokolov2011,javanainen2014,javanainen2016,cherroret2016}. Those approaches aim to go beyond the mean-field model developed earlier by Friedberg and co-authors~\cite{friedberg1973}. Indeed, while the mean-field model successfully predicted the Lorentz-Lorenz shift and the cooperative Lamb
shift observed in a thin atomic vapor cell~\cite{keaveney2012}, it fails to explain the observations in cold atomic systems~\cite{jennewein2016,bromley2016,jennewein2018}. It seems, indeed that dipole-dipole interaction, which is not considered in the mean field approach, become a dominant mechanism when Doppler broadening is absent~\cite{javanainen2014, jenkins2016, zhu2016}. Numerical simulations of microscopical models, as coupled dipole equations, are now commonly used to address those problems~\cite{javanainen1999,javanainen2014,jenkins2016,jennewein2016,corman2017}. They are useful in taking account of the sizes and shapes of the atomic clouds encountered experimentally. Those numerical methods are usually in fair agreement with experiments, but unfortunately do not always give a clear understanding of the basic physical mechanisms at play.

In this paper, we develop a perturbative model of the coherent transmission of light through an atomic medium where the scatterers are classical particle and considered at rest. Using configuration averaging in a slab geometry, we calculate the relative permittivity for an atomic medium at zero temperature, up to $1/k_0^2\ell_0^2$, by expanding the self-energy operator of light scattering in scattering diagrams. Here, $k_0$ is the resonant wave vector of the transition, and $\ell_0$ is the resonant ISA mean free path. Even though the perturbative method is limited to a density that is not too large, it gives analytical expressions with clear physical origins of the modifications to the refractive index. In particular, recurrent scattering of light~\cite{vantiggelen1994,wiersma1995,pierrat2010,aubry2014} (which arises from dipole-dipole interactions) and position correlation of the atoms~\cite{frisch1968} are two main physical mechanisms that modify the refractive index of an atomic cloud. Similar approaches were done in the past for quantum gas~\cite{morice1995,ruostekoski1997, ruostekoski1999}. Our theoretical results are in agreement with those previous works when they are taken at the classical limit. In addition, our perturbative expansion method allows an extension to cases where the atoms are moving~\cite{pierrat2009}. This latter point might be of particular importance to understand how the temperature acts as a dephasing mechanism in a collective scattering regime.

We compare the theoretical results to coupled dipole simulations of light transmission through the atomic medium.   In principle, the coupled dipole simulations can be performed in the dense regime. However, a large number of atoms is required to correctly simulate the bulk behavior, which requires a lot of computational resources and time. We are limited to a highest number density that corresponds to $k_0\ell_0=9.1$. Although this is still far from the dense regime of $k_0\ell_0\sim1$, modifications to the ISA can already be observed.

This article is organized as follows. In Sec.~\ref{sec.theory}, we present the theoretical formulation of the relative permittivity up to the second order term proportional to $1/k_0^2\ell_0^2$. We introduce the scalar
wave formulation -- neglecting the near field and light polarization -- before going to the more complex vector wave formulation. We discuss in detail the three contributions: ISA, pairwise positions correlation and recurrent scattering in both formulations. Some technical aspects of the calculation are given in the appendices. In Sec.~\ref{sec.numerical}, the results are compared to a numerical simulation of the transmission of light using the coupled dipole equations. We find good agreement with the perturbative calculations as far as the quantity $k_0\ell_0$ remains larger than unity.

\section{Theoretical formulation} \label{sec.theory}
% ===============================

We consider a large scattering medium of volume $V$ containing randomly-positioned motionless atomic scatterers. The
number of atoms inside the medium is denoted by $N_a$, with $N_a \gg 1$. We assume a uniform distribution for the
position of the scatterers, where the single-scatterer probability density to find a scatterer at $\bm{r}_j$ is given by
$P_1(\bm{r}_j) = 1/V$.

In general, each atomic scatterer carries an exclusion volume around it, \ie, a second scatterer cannot be found within a
distance less than $d_\text{min}$ from the first one. This assumption is useful for the numerical simulations, as it
avoids possible divergences. Finally, the limit
$d_\text{min}\rightarrow 0$ may be taken to describe the experimental results.

The two-scatterer probability density for a scatterer at $\bm{r}_p$ and a second one at $\bm{r}_q$
is~\cite{frisch1968}
\begin{equation}
   P_2(\bm{r}_p,\bm{r}_q) = P_1(\bm{r}_p)P_1(\bm{r}_q)\Big[1+h(\bm{r}_p,\bm{r}_q)\Big],
\end{equation}
where the function $h(\bm{r}_p,\bm{r}_q)$ is the pair correlation function between
the $p$-th and $q$-th scatterers. For a statistically homogeneous and isotropic medium, the pair correlation
function depends only on the separation $|\bm{r}_p-\bm{r}_q|$ of the two scatterers. For ``hard sphere'' atoms with the
exclusion volume, $h(d)$ is a complicated function of the inter-particle distance $d$~\cite{percus1958, wertheim1963}.
However, at a sufficiently low scatterer density, such as the one we consider here, we can approximate:
\begin{equation}
   h(d) =
   \begin{cases}
      -1 & \text{if $d < d_{\text{min}}$}\\
      0 & \text{otherwise}.
   \end{cases}
   \label{eq:paircorrfunc}
\end{equation}
The formulation discussed in the following actually works for an arbitrary pair correlation function, including those describing the quantum statistics of Bose and Fermi gases. It allows experimental studies of the effect of pair correlation function, as has been demonstrated in Ref.~\onlinecite{bons2016}. 

The atomic scatterers are treated as two-level atoms with a resonance frequency $\omega_0$. There is no absorption of light in the medium; all of the energy is elastically rescattered by the atoms. We consider only the case where the
intensity of the incident wave is much smaller than the saturation intensity of the transition, discounting any
nonlinear effect. The polarizability of the scatterers, in the rotating wave approximation~\cite{allen1974}, is
given by
\begin{equation}
   \alpha = -\frac{\alpha_0}{2} \frac{\Gamma}{\omega-\omega_0 + i\Gamma/2},
\end{equation}
where $k_0=\omega_0/c$. For scalar waves, $\alpha_0 = 4\pi/k_0^3$. For vector waves, $\alpha_0=6\pi/k_0^3$.

The system is illuminated by a monochromatic plane wave at frequency $\omega_L$, with a wave-vector $\bm{k}_L$. The
detuning of the incident wave is $\delta=\omega_L-\omega_0\ll \omega_0$.

\subsection{Scalar waves}
% ----------------------

In the scalar wave formulation, we disregard the
light polarization and describe the wave by a scalar ``electric field''.  In the frequency domain, the incident electric field
at  position $\bm{r}$ is denoted by
\begin{equation}
   E_{\textrm{in}}(\bm{r}) = E_0 \exp(- i\bm{k}_L\cdot\bm{r})
\end{equation}
where $E_0$ is the amplitude.

The electric field at any position $\bm{r}$ is given by the coherent superposition between the
incident field and the field radiated by all the atomic dipoles:
\begin{equation}
   E(\bm{r}) = E_{\textrm{in}}(\bm{r}) + \mu_0\omega_L^2\sum_{i=1}^{N_a} G_0(\bm{r}-\bm{r}_i)
   p(\bm{r}_i)\label{eq:flfield_prev}
\end{equation}
where $\mu_0$ is the vacuum permeability, $p(\bm{r}_i)$ the dipole moment of an atom at $\mathbf{r}_i$ and $G_0(\bm{r}-\bm{r}_i)$ the
free space scalar Green function that connects a point source dipole to its radiated field. It is given by
\begin{equation}
   G_0(\bm{r}-\bm{r}') = \frac{1}{4\pi |\bm{r}-\bm{r}'|} \exp(i k_L |\bm{r}-\bm{r}'|),
   \label{eq:G0}
\end{equation}
where $k_L = |\bm{k}_L|$.  In the literature~\cite{akkermans2007}, the Green function is sometimes defined with a minus sign
compared to Eq.~(\ref{eq:G0}). As a consequence, the self-energy computed later is also modified by a minus sign. The
sign convention has of course, no consequence, for physically measurable quantities such as the permittivity.  The dipole
$p(\bm{r}_i)$ induced on atom $i$ is given by
\begin{equation}
   p(\bm{r}_i)=\epsilon_0\alpha E_\text{ex}(\bm{r}_i)\label{eq:dipole}
\end{equation}
where $\epsilon_0$ is the vacuum permittivity and $E_\text{ex}(\bm{r}_i)$ the field exciting the atom (\ie the field
shining on the atom). It is given by the coherent superposition between the incident field and the field radiated by all other
atoms:
\begin{equation}
   E_\text{ex}(\bm{r}_i) = E_{\textrm{in}}(\bm{r}_i) + \mu_0\omega_L^2\sum_{\substack{l=1\\l\neq i}}^{N_a}
   G_0(\bm{r}_i-\bm{r}_l)p(\bm{r}_l).\label{eq:flexc_prev}
\end{equation}
Thus, combining \eq{flfield_prev} and \eq{flexc_prev} with \eq{dipole} leads to a set of equations, which allows us to compute
the electric field at any position
\begin{align}
   E(\bm{r}) & = E_{\textrm{in}}(\bm{r}) + \alpha k_L^2\sum_{i=1}^{N_a} G_0(\bm{r}-\bm{r}_i)
   E_\text{ex}(\bm{r}_i)\label{eq:flfield},
\\
   E_\text{ex}(\bm{r}_i) & = E_{\textrm{in}}(\bm{r}_i) + \alpha k_L^2\sum_{\substack{l=1\\l\neq i}}^{N_a}
   G_0(\bm{r}_i-\bm{r}_l) E_\text{ex}(\bm{r}_l).\label{eq:flexc}
\end{align}

The coherent electric field, in the forward direction, is given by the average electric field denoted by $\langle E(\bm{r})\rangle$. This average
is computed over all the possible configurations of the positions of the atomic scatterers. In practice, this ensemble
average is carried out by averaging the positions of the scatterers according to their probability distribution.
Experimentally, for a cold atomic cloud, the average is performed by a time integration of the signal collected in the forward direction by a
CCD camera, with a small numerical aperture.

The coherent field obeys the following equation known as the Dyson equation~\cite{dyson1949a,dyson1949b},
\begin{equation}
   \langle E(\bm{r}) \rangle = E_{\textrm{in}}(\bm{r}) + \iint G_0(\bm{r}-\bm{r}') \mstot(\bm{r}'-\bm{r}'')
      \langle E(\bm{r}'')\rangle \mathrm{d}^3\bm{r}'\mathrm{d}^3\bm{r}''.\label{eq:dyson}
\end{equation}
where $\mstot(\bm{r}'-\bm{r}'')$ is the electromagnetic wave analogue of the self-energy for the
scattering of quantum particles~\cite{akkermans2007}:
The self-energy here contains all scattering processes between a scatterer at $\bm{r}'$ and another at $\bm{r}''$, that
cannot be broken up into two or more independent scattering processes. In a statistically homogeneous medium, the Green function in the Fourier space obeys the following equation:
\begin{equation}
   \langle G\rangle(\bm{k}) = G_0(\bm{k}) + G_0(\bm{k}) \mstot(\bm{k}) \langle G\rangle(\bm{k}),\label{eq:gdyson}
\end{equation}
where $\langle G\rangle$ is the average Green function of light in the atomic medium, a diagonal operator in
$\bm{k}$-space.  The free space Green function in $\bm{k}$-space is given by the following Fourier
transform,
\begin{equation}
   G_0(\bm{k}) =\int G_0(\bm{r}') \exp(-i\bm{k}\cdot\bm{r}') \mathrm{d}^3\bm{r}'=\frac{1}{k^2 - k_L^2}. \label{eq:fsgreenk}
\end{equation}
Thus, from~\eq{gdyson}, the average Green function is given by
\begin{equation}
   \langle G\rangle(\bm{k}) = \frac{1}{k^2 - k_L^2-\mstot(\bm{k})}.\label{eq:greeneff1}
\end{equation}
For a statistically homogeneous medium, we expect that the average Green function takes the same form as~\eq{fsgreenk}, that is
\begin{equation}
   \langle G\rangle(\bm{k}) = \frac{1}{k^2 - k_\text{eff}^2},\label{eq:greeneff2}
\end{equation}
where $k_\text{eff}$ is the effective wavevector associated to the relative permittivity
$\epsilon_\text{r}=k_\text{eff}^2/k_L^2$. Comparing~\eqs{greeneff1}{greeneff2}, we have $k_\text{eff}^2 = k_L^2 +
\mstot(\bm{k})$. In the most general case, the effective wavevector is non-local (\ie, its magnitude depends on
$\bm{k}$). Nevertheless, if $\mstot(\bm{k})\ll k_L^2$ which is usually the case, the average Green
function is very peaked around $k_L$ and the self-energy $\mstot(\bm{k})$ can be approximated by $\mstot(\bm{k}_L)$.
This is the so-called \emph{on-shell} approximation. Therefore, the  relative permittivity
is~\cite{lagendijk1996},
\begin{equation}
    \epsilon_\text{r} \equiv \left(\frac{k_\text{eff}}{k_L}\right)^2 = 1 + \frac{\mstot(k_L)}{k_L^2}.
\end{equation}

Note that, in general, the relative permittivity is a complex quantitiy. The index of
	refraction, given by
\begin{equation}
n = \sqrt{\epsilon_\mathrm{r}} \approx  1 + \frac{\mstot(k_L)}{2k_L^2}
\label{eq:refractive_index}
\end{equation}
is also complex. Its imaginary part describes the exponential attenuation of the coherent
beam through the disordered medium thanks to scattering.

\begin{figure}[h]
   \centering
   \includegraphics[width = \linewidth]{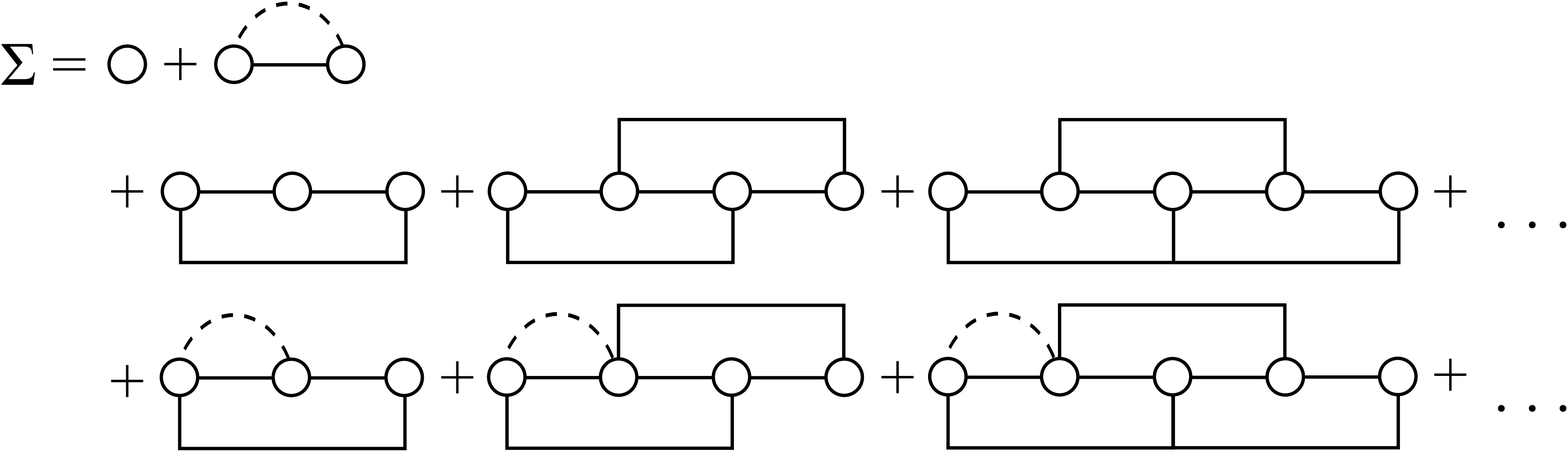}
   \caption{\label{fig:diagram} The self-energy expanded in the diagrammatic representation method. The first
   line consists of the diagrams representing ISA and pairwise correlation in the scatterer positions. The second line
   consists of diagrams that arise from pure recurrent scattering of varying orders. The third line consists of diagrams
   that include contributions from position correlation and recurrent scattering. Each
   diagram here contains an average over the positions of all the atomic
   scatterers. See the text for the meanings of the open circles and lines in these scatterings diagrams.}
\end{figure}

From the above equations, the calculation of the self-energy is needed to find an expression for
$\epsilon_\text{r}$.  In general, computing the exact form of the self-energy is very complicated, but perturbative
diagrammatic methods exist, which expand the self-energy in special kinds of diagrams that represent the scattering
processes in the medium. Additional details of this approach can be found in numerous references, for example in
Refs.~\onlinecite{frisch1968,rytov1989,lagendijk1996}. The self-energy is written as a sum of irreducible
diagrams, namely those that cannot be separated into two sub-diagrams by cutting one of the lines in the
diagram. Position averaging is implied in these diagrams. In~\fig{diagram}, we write out all the
contributing diagrams up to the order of $1/k_0^2\ell_0^2$ for the self-energy. The open circles in the diagrams
represent scattering events. There are two types of solid lines in the diagram. Those that join adjacent scattering
events represent the propagation of the wave between two scattering events; the other solid lines join two scattering
events that occur at the same scatterer. Finally, the dashed lines between two scatterers indicate that they are
correlated in their positions. Here, this is due to the exclusion volume around each atom.

The lowest order diagram in \fig{diagram} consists of just one scattering event.
It describes the situation where each atomic scatterer scatters light independently of one another. This is the ISA
contribution. The second term contains two scattering events involving distinct scatterers that are correlated in their
positions. This gives a second order contribution to the self-energy. The diagrams in the second line of~\fig{diagram}
give all the contributions in second order from pure recurrent scattering between two scatterers. The diagrams in
the third line includes in recurrent scattering, the effect of correlation in scatterer positions. For convenience, we
separate the self-energy into three terms,
\begin{equation}
   \mstot = \msisa + \mscor + \msrec,
\end{equation}
where $\msisa$ is the first order term that gives us the ISA, $\mscor$ is the second order term with pair correlation in the
positions (the second diagram in first line of~\fig{diagram}) and $\msrec$ includes all the remaining diagrams arising
from recurrent scattering. We discuss each of these contributions separately in the following.

\subsubsection{Independent scattering approximation}
% ..................................................

We calculate now the first diagram in~\fig{diagram}, which is the ISA contribution. The
self-energy is computed as follows:
\begin{equation}
   \msisa(\bm{r}-\bm{r}') = \sum_{i=1}^{N_a} \int \alpha k_L^2 \delta(\bm{r}-\bm{r}_1)
      P_1(\bm{r}_1)\mathrm{d}^3\bm{r}_1 \delta(\bm{r}-\bm{r}').\label{eq:ISAdiag}
\end{equation}
Evaluating the above integral, we find in the $\mathbf{k}$-space,
\begin{equation}
   \frac{\msisa}{k_L^2} = \rho\alpha.\label{eq:ISAsigma}
\end{equation}
Here, we have denoted the scatterers density as $\rho = N_a/V$. Thus, in the ISA regime, the relative
permittivity is
\begin{equation}
   \epsilon_\text{\tiny{ISA}} = 1 + \rho \alpha. \label{eq:ISA}
\end{equation}
It is also possible to write the above expression in terms of the strength of the disorder $k_0\ell_0$. For this
purpose, we can rewrite the two-level atomic polarizability as $\alpha = \alpha_0 \tilde{\alpha}$, where
$\tilde{\alpha}$ contains the $\delta$ dependence of $\alpha$:
\begin{equation}
\tilde{\alpha} = -\frac{\Gamma/2}{\omega-\omega_0+i\Gamma/2} = \frac{i}{1-2i\delta/\Gamma}
\end{equation}

The scattering cross section at resonance is given by $\sigma_s = k_0 \alpha_0$. Additionally, the mean free path at
resonance is given by $\ell_0 = 1/(\rho\sigma_s)$. We finally obtain $\rho\alpha_0 = 1/(k_0\ell_0)$, connecting the
density to $k_0\ell_0$. Hence, the contribution in the ISA regime is first order in $1/k_0\ell_0$ (or equivalently the
first order in $\rho$),
\begin{equation}
   \epsilon_\text{\tiny{ISA}} = 1 + \frac{1}{k_0\ell_0}\tilde{\alpha}. \label{eq:epsilonISA}
\end{equation}

\subsubsection{Position correlations}
% ....................................

The other diagrams in~\fig{diagram} are second order diagrams. The second diagram  is
related to the position correlation between pairs of atoms. Its contribution to the self-energy is calculated to be
\begin{multline}
   \mscor(\bm{r}-\bm{r}') = \sum_{i=1}^{N_a}\sum_{\substack{j=1\\j\neq i}}^{N_a}
      \iint \alpha k_L^2 \delta(\bm{r}-\bm{r}_1) G_0(\bm{r}_1-\bm{r}_2)
\\
   \times \alpha k_L^2 \delta(\bm{r}_2-\bm{r}')P_1(\bm{r}_1) P_1(\bm{r}_2) h(|\bm{r}_1-\bm{r}_2|)
      \mathrm{d}^3\bm{r}_1\mathrm{d}^3\bm{r}_2.\label{eq:scorr}
\end{multline}
We use the pair correlation function
of~\eq{paircorrfunc} to evaluate the integral. We then perform the Fourier transform to finally arrive at the following equation in
$\bm{k}$-space:
\begin{equation}
   \frac{\mscor}{k_L^2} = -\frac{1}{(k_0\ell_0)^2}\tilde{\alpha}^2 \left[\frac{1}{4}\left(1-e^{2i\tilde{d}_\text{min}}\right)
      +\frac{i}{2}\tilde{d}_\text{min}\right]\label{eq:scorrfinal}
\end{equation}
where $\tilde{d}_\text{min} = k_L d_\text{min}$. We have made the \emph{on-shell} approximation $k\approx k_L$, with
$k=|\bm{k}|$.   The
details of this derivation is found in Appendix~\ref{app.scalar1}. We note from~\eq{scorrfinal} that in the limit of
$d_\text{min}\rightarrow 0$, $\mscor$ goes to zero, as expected in the scalar approximation.

\subsubsection{Recurrent scattering}
% ..................................

The two-atom recurrent scattering diagrams in the second and third lines of \fig{diagram}, come in different orders with
varying number of scattering events. For example, the first diagrams in the second and third lines contain three
scattering events. This is the simplest possible recurrent scattering between two distinct scatterers.  We will denote
the sum of these two diagrams as $\msrec^{(n)}$. The number in the superscript denotes the number of scattering events,
$n=3$ in this case. Now, we sum over all orders,
\begin{equation}
   \msrec = \sum_{n=3}^\infty \msrec^{(n)},
\end{equation}
with each $\msrec^{(n)}$ terms containing one diagram from the second line and one diagram from the third line
of~\fig{diagram}.  We can further distinguish two different types of recurrent scattering diagrams. The first type
consists of diagrams where the first and last scattering events happen at the same scatterer. All $\msrec^{(n)}$ terms
with an odd value of $n$ falls under this category. These diagrams are known as the loop diagrams (see
Ref.~\cite{vantiggelen1994}). The simplest example is $\msrec^{(3)}$:
\begin{multline}
   \msrec^{(3)}(\bm{r}-\bm{r}') = \sum_{i=1}^{N_a}\sum_{\substack{j=1\\j\neq i}}^{N_a} \iiint \alpha k_L^2 \delta(\bm{r}-\bm{r}_1) \\
   \times G_0(\bm{r}_1-\bm{r}_2) \alpha k_L^2 G_0(\bm{r}_2-\bm{r}_3)\alpha k_L^2\delta(\bm{r}_3-\bm{r}')\\
   \times P_2(\bm{r}_1,\bm{r}_2) \mathrm{d}^3\bm{r}_1\mathrm{d}^3\bm{r}_2\mathrm{d}^3\bm{r}_3\delta(\bm{r}-\bm{r}').\label{eq:srec3}
\end{multline}
The second type of diagrams have the first and last scattering events occurring at two different scatterers; they
consists of $\msrec^{(n)}$ terms with even value of $n$. They are classified as the boomerang diagrams in
Ref.~\cite{vantiggelen1994}. The simplest example is $\msrec^{(4)}$:
\begin{multline}
   \msrec^{(4)}(\bm{r}-\bm{r}') = \sum_{i=1}^{N_a}\sum_{\substack{j=1\\j\neq i}}^{N_a}
      \idotsint \alpha k_L^2 \delta(\bm{r}-\bm{r}_1) G_0(\bm{r}_1-\bm{r}_2) \\
   \times \alpha k_L^2G_0(\bm{r}_2-\bm{r}_3)\alpha k_L^2G_0(\bm{r}_3-\bm{r}_4)\alpha k_L^2\delta(\bm{r}_4-\bm{r}')\\
   \times P_2(\bm{r}_1,\bm{r}_2)\delta(\bm{r}-\bm{r}_3)\delta(\bm{r}_2-\bm{r}')
      \mathrm{d}^3\bm{r}_1\dots\mathrm{d}^3\bm{r}_4.\label{eq:srec4}
\end{multline}
Summing up all the diagrams with recurrent scattering, we obtain in $\bm{k}$-space,
\begin{multline}
   \msrec(\bm{k})= (\rho\alpha_0^2) k_L^4 \sum_{l=0}^{\infty} \left(\alpha_0 k_L^2\right)^{2l+1}\tilde{\alpha}^{2l+3}\\
   \times\Bigg[\int_{V-V_\text{ex}} G_0^{2l+2}(\bm{R}')\mathrm{d}^3\bm{R}'+\left(\alpha_0 k_L^2\right)\tilde{\alpha}\\
   \times\int_{V-V_\text{ex}} G_0^{2l+3}(\bm{R}')e^{-i\bm{k}\cdot\bm{R}'}\mathrm{d}^3\bm{R}'\Bigg]\label{eq:totalrec}
\end{multline}
where $V_\text{ex}$ represents the exclusion volume.  There are two integrals inside the summation. The first integral
is associated with the loop diagrams, while the second integral is associated with boomerang diagrams. The
summations clearly have the structure of geometric series, making it possible to resum the infinite number of terms, see
Appendix B for the details of the calculation. Similarly to the case of $\mscor$, we approximate $k\approx k_L$.
Evaluating the above integral results in
\begin{equation}
   \frac{\msrec}{k_L^2} = \frac{1}{(k_0\ell_0)^2} \frac{i\tilde{\alpha}^3}{2}\left(e^{2i\tilde{d}_\text{min}} + \tilde{\alpha} I_s\right),
   \label{eq:recscalar}
\end{equation}
with
\begin{equation}
   I_s = \bigintssss_{\tilde{d}_\text{min}}^\infty \frac{e^{2ix}\left(1-\left(1+2i\tilde{\alpha}\right) e^{2ix}\right)}{x^2-\tilde{\alpha}^2e^{2ix}}\mathrm{d}x.
\end{equation}
The integral $I_s$ is computed numerically using a cut-off on the upper limit of the integral. We found that a cut-off at $x=80000$ is sufficient for the integral to converge. The
numerical integration is performed using an adaptive algorithm for oscillating integrand.

\subsection{Vector waves}
% -----------------------

For the case of vector waves, the incident field in the frequency domain is given by
\begin{equation}
   \bm{E}_{\textrm{in}}(\bm{r}) = \bm{E}_0 \exp(-i\bm{k}_L\cdot\bm{r}).
\end{equation}
The free space Green function is now  a dyadic given by~\cite{tai1993}
\begin{multline}
   \tensor{\bm{G}}_0(\bm{r}-\bm{r}') = \frac{k_L}{4\pi}\Bigg[\beta(k_L|\bm{r}-\bm{r}'|) \tensor{\bm{P}} \\
      +\gamma(k_L|\bm{r}-\bm{r}'|) \tensor{\bm{U}}\Bigg]- \frac{\tensor{\bm{I}}}{3k_L^2}\delta (\bm{r}-\bm{r}'),
      \label{eq:G0_vec}
\end{multline}
where
\begin{align*}
\tensor{\bm{U}} & = \bm{u}\otimes\bm{u}\quad\text{with }\bm{u}=(\bm{r}-\bm{r}')/|\bm{r}-\bm{r}'|,
\\
  \tensor{\bm{P}} & = \tensor{\bm{I}} - \tensor{\bm{U}},
\\
   \beta(x) & = \frac{e^{ix}}{x}\left(1 - \frac{1}{ix} - \frac{1}{x^2}\right),
\\
   \gamma(x) & = \frac{2e^{ix}}{x}\left(\frac{1}{ix} + \frac{1}{x^2}\right),
\end{align*}
and $\tensor{\bm{I}}$ is the identity dyadic.
In contrast with the scalar case where the Green function has a $1/r$ divergence at the
origin, the vector wave case contains near field effects with additional $1/r^2$ and $1/r^3$ singularities.

 In the Fourier space, the Green function is given by
\begin{equation}
   \tensor{\bm{G}}_0(\bm{k}) = \frac{1}{k^2-k_L^2}\tensor{\bm{P}}_k - \frac{1}{k_L^2}\tensor{\bm{K}}. \label{eq:vecGfree}
\end{equation}
The dyadic $\tensor{\bm{K}} = \bm{k}\otimes\bm{k} / k^2$, is the projector along the direction of $\bm{k}$. The dyadic
$\tensor{\bm{P}}_k = \tensor{\bm{I}} - \tensor{\bm{K}}$ projects onto the space orthogonal to $\bm{k}$. The
longitudinal part of the Green function (proportional to $\tensor{\bm{K}}$) does not propagate much further than one
wavelength. Thus, only the transverse component (proportional to $\tensor{\bm{P}}_k$) is relevant for the coherent
transmission of light through a medium much thicker than the wavelength. However, one must carefully keep the
full spatial dependence of the Green function -- including both the longitudinal and the transverse parts -- when computing the effect of
position correlations and recurrent scattering. The average Green function also splits into longitudinal
and transverse components:
\begin{equation}
   \langle\tensor{\bm{G}}\rangle(\bm{k}) = \frac{1}{k^2-k_L^2-\mttot(\bm{k})}\tensor{\bm{P}}_k - \frac{1}{k_L^2+\Sigma_l(\bm{k})}\tensor{\bm{K}},\label{eq:vecGave}
\end{equation}
where the dyadic self-energy $\tensor{\bm{\Sigma}}$ is also separated into
\begin{equation}
\mvtot = \Sigma_l \tensor{\bm{K}} + \mttot \tensor{\bm{P}}_k.
\end{equation}
Comparing the transverse components of \eqs{vecGfree}{vecGave}, which are relevant in the coherent transmission of light, we have
\begin{equation}
   \epsilon_\text{r} =  1 + \frac{\mttot}{k_L^2}.
\end{equation}
Thus, for a statistically homogeneous and isotropic scattering medium under consideration here, the relative
permittivity $\epsilon_\text{r}$ in the vector wave case, is still a scalar quantity.
The self-energy $\mvtot$, in the vector case, is also given by the diagrams in~\fig{diagram}, with the same
interpretation. Following the case of scalar waves, we separate the self-energy into three terms
representing the ISA contribution, the positions correlations contribution and the recurrent scattering contribution:
\begin{equation}
   \mvtot = \mvisa + \mvcor + \mvrec.
\end{equation}

\subsubsection{Independent scattering approximation}
% ..................................................

Since the diagrams and their interpretations are the same as in the scalar case, we can write an
integral for the ISA contribution of vector waves, similar to~\eq{ISAdiag}. We get
\begin{equation}
   \frac{\mvisa}{k_L^2}= \frac{1}{k_0\ell_0}\tilde{\alpha} \tensor{\bm{I}}, \quad \text{and} \quad \frac{\mtisa}{k_L^2} = \frac{1}{k_0\ell_0}\tilde\alpha,
\end{equation}
which is similar to~\eq{ISAsigma}. $\mtisa$ is the transverse component of $\mvisa$. The ISA relative permittivity is also given by~\eq{epsilonISA}.

\subsubsection{Position correlations}
% ....................................

For the position correlation contribution, we solve the dyadic version of~\eq{scorr}, using the dyadic Green function
for vector waves. The details of the calculations are given in Appendix~\ref{app.vector1}. The resulting expression of the transverse component $\mtcor$ is
\begin{equation}
   \frac{\mtcor}{k_L^2}= \left(\frac{1}{k_0\ell_0}\right)^2\tilde{\alpha}^2 \mathcal{C}\left({\tilde{d}_\text{min}}\right),\label{eq:posc}
\end{equation}
where
\begin{multline}
   \mathcal{C}\left({\tilde{d}_\text{min}}\right) =\frac{2i+2i{\tilde{d}_\text{min}}^2-{\tilde{d}}_\text{min}^3-2i{\tilde{d}_\text{min}}^4}{4{\tilde{d}_\text{min}}^3}\\
   -\frac{\left(2i+4{\tilde{d}_\text{min}}-2i{\tilde{d}_\text{min}}^2-{\tilde{d}_\text{min}}^3\right)\mathrm{e}^{2i{\tilde{d}_\text{min}}}}{4{\tilde{d}_\text{min}}^3}.
   \label{eq:posc2}
\end{multline}
It is interesting to note that $\mtcor / k_L^2$ is nonzero when $d_\text{min}\rightarrow0$, unlike in the case of scalar waves. In fact,
\begin{equation}
   \frac{\mtcor}{k_L^2} = \frac{1}{(k_0\ell_0)^2}\frac{\tilde{\alpha}^2}{3},
\end{equation}
when $d_\text{min}\rightarrow0$. This is due to the $\delta(\bm{r}-\bm{r}')$ term in the free Green function in \eq{G0_vec} being absent in the scalar case. This term is responsible for the so-called Lorentz-Lorenz
shift~\cite{mallet2005,guerin2006}.

\subsubsection{Recurrent scattering}
% ..................................

Similarly to the case of scalar waves, the following series has to be summed for the recurrent scattering contribution,
\begin{equation}
   \mvrec = \sum_{n=3}^\infty \mvrec^{(n)},
\end{equation}
where $\mvrec^{(n)}$ is computed from the two recurrent scattering diagrams having $n$ scattering events. They can be
expressed in the same way as in~\eqs{srec3}{srec4}, with the dyadic Green function being used instead. This leads to a
dyadic version of~\eq{totalrec}. The transverse component of $\mvrec$ is found to be:
\begin{equation}
   \frac{\mtrec}{k_L^2}= \frac{1}{(k_0\ell_0)^2}\frac{\tilde{\alpha}^3}{2} I_v,\label{eq:vrec1}\\
\end{equation}
where
\begin{multline}
   I_v =  \int_{\tilde{d}_\text{min}}^\infty \frac{2x^2\left[\beta(x)^2 + \frac{9}{4}\tilde{\alpha}(\omega)\beta(x)^3\left\{j_0(x)-j_1(x)/x\right\}\right]}{1-\frac{9}{4}\tilde{\alpha}(\omega)^2 \beta(x)^2}
   \\+x^2\frac{\gamma(x)^2 + \frac{9}{2}\tilde{\alpha}(\omega)\gamma(x)^3j_1(x)/x}{1-\frac{9}{4}\tilde{\alpha}(\omega)^2 \gamma(x)^2}\,\mathrm{d} x.
   \label{eq:vrec2}
\end{multline}
$j_0(x)$ and $j_1(x)$ are the zeroth- and first-order spherical Bessel functions. The integral $I_v$ is evaluated
numerically. Details of this calculation are given in Appendix~\ref{app.vector2}.

\section{Numerical studies} \label{sec.numerical}
% =========================

\subsection{Setup for the numerical studies}
% ------------------------------------------

To check the validity of the theoretical expressions, a numerical study is carried out to simulate the coherent
transmission of light through a slab of atomic medium.  The relative
permittivity of the medium is extracted from the transmitted field. In the following, we discuss the coupled
dipole method that is used to perform the simulation. We discuss in detail the case of scalar waves.
The same method is applicable to the vector waves, with just a few differences.  These differences are pointed out as we
encounter them.

\begin{figure}[h]
   \centering
   \includegraphics[width = \linewidth]{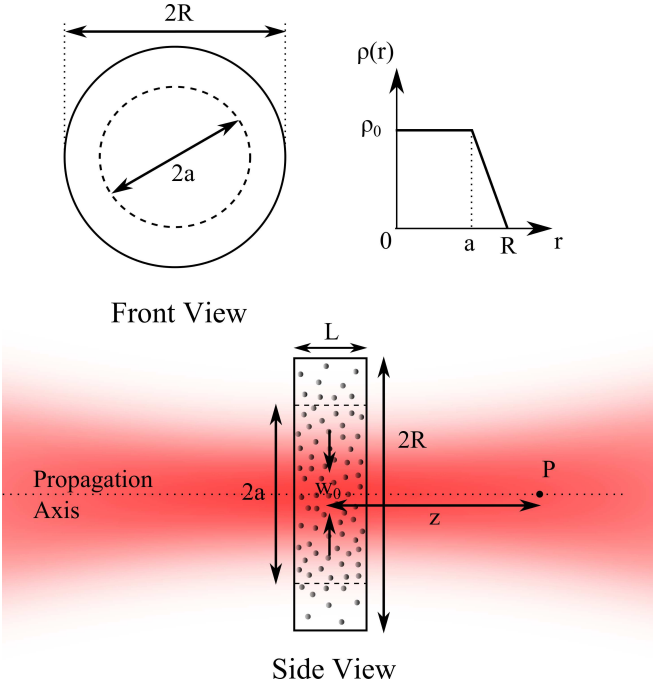}
   \caption{\label{fig:system}  Setup
   for the system  studied numerically. It consists of a cylinder of thickness $L$ under normal incidence of a
   Gaussian beam laser (waist $w_0$). The probe beam is focused at the geometric center of the
   cylinder. The density of the cylinder is distributed symmetrically around the axis, with two distinct regions. Within
   a radius $a$, the density of scatterers has a uniform value of $\rho_0$. Outside this region, the density
   decreases linearly until it becomes zero at the outer edge. The radial distribution of the density is plotted in the
   top right figure. The scatterers are motionless and are distributed randomly inside the cylinder according to this
   density distribution. We keep an exclusion volume of a sphere of radius $d_\mathrm{min}$ around each
   scatterer. The coherently transmitted field is computed on the propagation axis through the cylinder, at a distance
   $z$ away from the origin, which is at the center of the cylinder. The position $P$, where the coherent field is computed, ranges from a few
   wavelengths beyond the outgoing surface to 500 wavelengths away.}
\end{figure}

The atomic scatterers in our simulations are distributed randomly within a cylinder of thickness $L$ (see~\fig{system}) at fixed positions. The radius $R$ of the cylinder is chosen such that it is larger
than the thickness $L$, thereby making sure that the geometry is as close as possible to a slab.
The scatterers are distributed uniformly with density $\rho_0$ within a diameter of $2a$ in the plane perpendicular to
the propagation axis. Beyond this distance, the density of the scatterers linearly decreases until it becomes zero when
it reaches the edge of the cylinder. Configurations where any two scatterers are separated by less than
$d_\mathrm{min}$ are rejected. Within the range of parameters that we consider, we check that the approximate expression of $h$ in \eq{paircorrfunc} is valid.

The incident Gaussian beam has a waist of $w_0$. The value of the waist is chosen to be
smaller than the lateral dimension of the cylinder. The linearly decreasing density close to the edges of the cylinder,
reduces the diffraction of the incident beam at the edges. This also ensures that, along the propagation axis, we are
always in the shadow of the cylinder. However, the waist is large enough such that its Rayleigh length, $z_R =
\pi w_0^2/\lambda_L$ is larger than $L$, where $\lambda_L = 2\pi / k_L$ is the laser wavelength. Thus, we have a
well-defined direction of the wave vector along $\bm{k}_L$ inside the cylinder. The incident beam is focused at the
geometrical center of cylinder, which is chosen as the origin of our coordinate system. The incident electric field
for the numerical studies is given by
\begin{multline}
   E_\text{in}(\bm{r}) = E_0(z)\exp(i\bm{k}_L\cdot \bm{R})\exp\left[- \frac{x^2+y^2}{w_0^2(1+iz/z_R)}\right],\label{eq:Ez}
\end{multline}
where $z=\bm{k}_L\cdot\bm{r}/k_L$ is the distance along the propagation axis, and $x,y$ are the transverse coordinates. $E_0(z)$ is the amplitude of the field
along the central propagation axis, and is given by
\begin{equation}
   E_0(z) = \frac{1}{1+iz/z_R}.
\end{equation}
For vector waves,  the polarizations of the coherently transmitted beam and the incoming beam are identical, and can
be disregarded. Therefore, \eq{Ez} remains true for the vector waves. For the simulation of the vector waves, we choose for simplicity a linear polarization for the incident field.
%\begin{equation}
%   E_0(z) = \frac{1}{1+iz/z_R}.
%\end{equation}
In general, the field can be computed at any arbitrary position, however, we only calculate the field on the propagation
axis ($x=0$ and $y=0$). The incident field amplitude $E_0(z)$ is used to normalize the coherently transmitted field $\langle E(z)\rangle$,
at a distance $z$ along the propagation axis.

Our theoretical calculation of the relative permittivity,  assumed a
statistically translational invariant system. This approach is not strictly valid near an interface where the density varies abruptly. The depth of the
skin layer is typically of the order of $1/k_L$. If the thickness of the medium is sufficiently large -- such that
$k_LL\gg 1$ -- the index of refraction can be taken inside the medium as if it was infinite. Consequently, the average
field in the medium varies like $\exp(i k_L n z)$, where the refractive index is given by
Eq.~(\ref{eq:refractive_index}), $n=\sqrt{\epsilon_\text{r}}$. Because the refractive  index is not unity, there is an
index mismatch both at the ingoing and outgoing interfaces. This leads to  partial reflections of the incoming beam, which
can be calculated using standard formula~\cite{born1999, javanainen2016}. The transmitted field from a slab of thickness $L$ becomes
\begin{align}
   \langle E(z) \rangle/E_0(z) =& \frac{4n\exp[ik_L(n-1)L]}{(n+1)^2-(n-1)^2\exp(2ik_L n L)}\nonumber\\
   \equiv& F(n)\exp[ik_L(n-1)L].
\end{align}
From the above equation, it is not straightforward to obtain the value of $n$ from the numerically calculated value of $\langle E(z) \rangle$. Nevertheless, since we are in a regime where the perturbative expansion in terms of atomic density is valid, we can also expand $F(n)$ up to second order in the density, to find
\begin{multline}
   \langle E(z) \rangle/E_0(z) \\= \left[1+\frac{1}{16}\left(\frac{\tilde\alpha}{k_0\ell_0}\right)^2\left(\mathrm{e}^{2ikn_{\text{ISA}}L}-1\right)\right]\exp[ik_L(n-1)L], \label{eq:blnum}
\end{multline}
with $n_\text{ISA} = 1+\tilde\alpha/(k_0\ell_0)/2$, the refractive index in the ISA regime. The equation above can now be solved for $n$,  from which $\epsilon_r$ is obtained.

Strictly speaking, \eq{blnum} holds only for a slab under plane wave
illumination. Since the parameters for the numerical studies are chosen to approximate the case of plane
wave illumination on a medium with a slab geometry, \eq{blnum} can be applied in our studies.

Note also that, the number of atoms in the cylinder has to be large, ${N_a\gg1},$  since the theoretical
expressions in the previous section are obtained in this limit.

\subsection{Coupled dipole simulation}
% ---------------------------------

The coupled dipole simulation is carried out by
solving~\eqs{flfield}{flexc} in the frequency domain. The corresponding vector equations are used when computing the
coherent transmission of vector waves.

The calculation is performed in two stages. The first stage consists in computing the external fields
$E_\text{ex}(\bm{r_j})$ for $j = 1,2,\dots,N_a$. This is achieved by solving~\eq{flexc}, which is a coupled linear
system with $N_a$ equations,  to find the $N_a$ values of $E_\text{ex}(\bm{r_j})$. In the case of vector waves, we have $3N_a$ equations involving $3N_a$ variables. Once all the values of
$E_\text{ex}(\bm{r}_j)$ are known, \eq{flfield} is used to compute $E(\bm{r})$. The total field $E(z)$ is computed on
the propagation axis through the cylinder by varying the value of $z$.  By far, the most CPU intensive stage is the
solution of the coupled linear equations, scaling like $N_a^3.$

\subsection{Coherent transmission}
% --------------------------------

\begin{figure}[h]
   \centering
   \includegraphics[width = \linewidth]{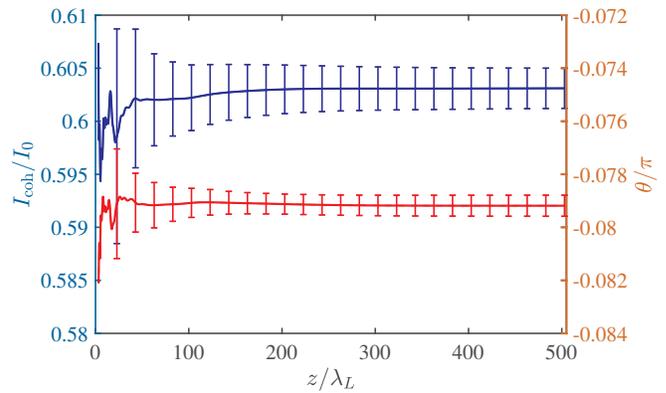}
   \caption{\label{fig:numres} Curves showing the position dependence (along the propagation axis) of the coherently
   transmitted intensity $I_\text{coh}$ (blue solid curve), and the phase difference $\theta$ between the coherent field and the
   incident field (red dashed curve). This example is shown for the vector wave case. The error bars are calculated for all points on the curves but only
   shown at selected points. The parameters in the calculation are $k_0\ell_0 = 36.4$ and $\delta
   = 0.5\Gamma$; see also the text for other computation parameters.}
\end{figure}

The calculation described above is repeated for different independent realizations of the scatterers positions inside
the cylinder.  The ensemble-averaged field $\langle E(z)\rangle$ at a point P that is sufficiently far away along the
scattering medium gives us the coherently transmitted field [see~\fig{numres} for one example of the position
dependence of the normalized coherent intensity $I_\text{coh}(z) = |\langle E(z) \rangle / E_0(z)|^2$ and phase
$\theta(z)$ of $\langle E(z) \rangle / E_0(z)$]. Close to the cylinder, at a distance comparable to the average
inter-atomic distance, the calculated field displays large statistical fluctuations.

\subsection{Computational errors}
% -------------------------------

In~\fig{numres}, we show the statistical error bars at few points on the curves. We now describe how the error associated with
each point is calculated. The number of configurations used to compute the coherent field is denoted as $N_\text{conf}$.
In order to compute the errors, we divide the number of realizations into $N_\text{part}$ partitions. The mean electric
field  $\langle E_p(z) \rangle$ is calculated for each partition. Here, $p$ is the index of the partitions. From the
$N_\text{conf}/N_\text{part}$ values of averaged fields computed for each partition, we compute their standard deviation
$\sigma_f$. Similarly, the phase shift of the transmitted field with respect to the incident field is also calculated
for each partition. The standard deviation is denoted as $\sigma_p$. The errors are then given by $\sigma_f /
\sqrt{N_\text{part}}$ and $\sigma_p / \sqrt{N_\text{part}}$, respectively, for the coherent field and the phase
difference. The error in the values of $\epsilon_\text{r}$ is then calculated by propagating the error accordingly. With large enough partitions, the error calculated is independent of $N_\text{part}$.

\subsection{Scalar waves}
% -----------------------

Using the numerical method described above, we study numerically the coherent light transmission at $k_0\ell_0 = 9.1$,
$18.2$ and $36.4$, for the case of scalar waves. In this study, we set $2R = \unit{35}{\micro\meter}$,
$2a=\unit{22}{\micro\meter}$, $w_0 = \unit{4.5}{\micro\meter}$, $N_\text{conf} = 3200$, $N_\text{part} = 40$,
$\tilde{d}_\text{min} = 0.455$ and the sample thickness $L$ is 1, 2 and 4~{\micro\meter} for the 3 cases studied. The corresponding values of the rescaled density $\rho_0/k_0^3$ is summarized in Table~\ref{tab.scalarvector}.
The wave vector used in the numerical simulation corresponds to the strontium $^1$S$_0\rightarrow^3$P$_1$
intercombination transition, \ie, $k_0=9.1\times10^{6}$~$\meter^{-1}$. This means that the scaled diameters of the
cylinder are $2k_0R=637$ and $2k_0a=400.4$, respectively. The scaled waist of the beam is $k_0w_0=41$. The number of atoms used in the
numerical simulation is $N_a= 4277$. With these parameters, we approximate as closely as possible a uniform slab of density $\rho_0$ used in the
theoretical study. Note that the
results depend only on the scaled parameters: changing the wavevector $k_0$ while keeping the same values of $k_0R,
k_0a, k_0w_0, k_0\ell_0$ produces exactly the same set of equations, and thus the same solutions.

\begingroup
   \begin{table}
      \caption{\label{tab.scalarvector} Values of $k_0\ell_0$ and $\rho_0/k_0^3$ for the cases studied
      numerically. The values are tabulated for both the scalar and vector waves. Here, we set $b_0=1$, and hence,
      $L=\ell_0$.}
      \begin{tabular}{c c c}
      \hline\hline
         $k_0\ell_0$ & $\rho_0/k_0^3$ (scalar) & $\rho_0/k_0^3$ (vector)\\
      \hline
         9.1 & $8.8\times10^{-3}$ & $5.8\times 10^{-3}$\\
         18.2 & $4.4\times10^{-3}$ & $2.9\times 10^{-3}$\\
         36.4 &  $2.2\times10^{-3}$ & $1.5\times 10^{-3}$\\
      \hline\hline
      \end{tabular}
   \end{table}
\endgroup

\begin{figure}[h]
   \centering
   \includegraphics[width = \linewidth]{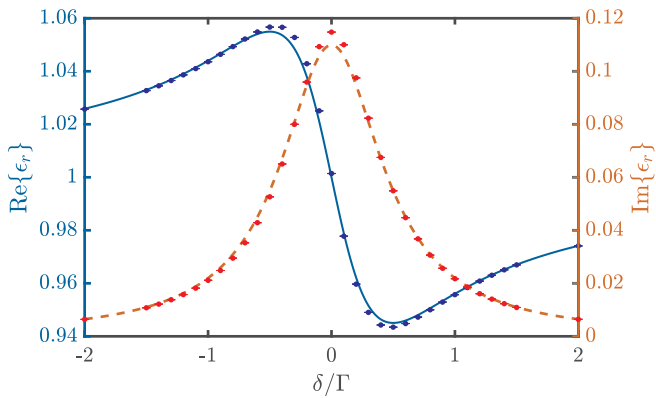}
   \caption{\label{fig:scalar1} Real and imaginary parts
   of the relative permittivity for scalar waves extracted from the numerical simulation results (blue and red dots
   for real and imaginary parts respectively).  The ISA prediction for the real part of $\epsilon_\text{r}$ is shown
   as the blue solid curve, while the imaginary part is shown as the red dashed curve. The numerical results are shown
   only for the case of $k_0\ell_0=9.1$, where deviations from the ISA prediction -- although obviously rather small -- are expected to be the largest among
   the $k_0\ell_0$ values computed.}
\end{figure}

The range of detuning computed in our study is $-3\Gamma \le \delta \le 3 \Gamma$. For each value of the detuning, we
compute the coherent field. The relative permittivity $\epsilon_\text{r}$ is calculated using~\eq{blnum}. We first
compare the numerical results to the ISA prediction for the case $k_0\ell_0 = 9.1$ (see~\fig{scalar1}). This is the
value where we expect the largest deviation from the ISA prediction. At first sight, the numerical results agree
very well with the ISA prediction. However, a close inspection shows there are indeed small deviations, particularly,
around $\delta = 0$.

\begin{figure}[h]
   \centering
   \includegraphics[width = 0.48\textwidth]{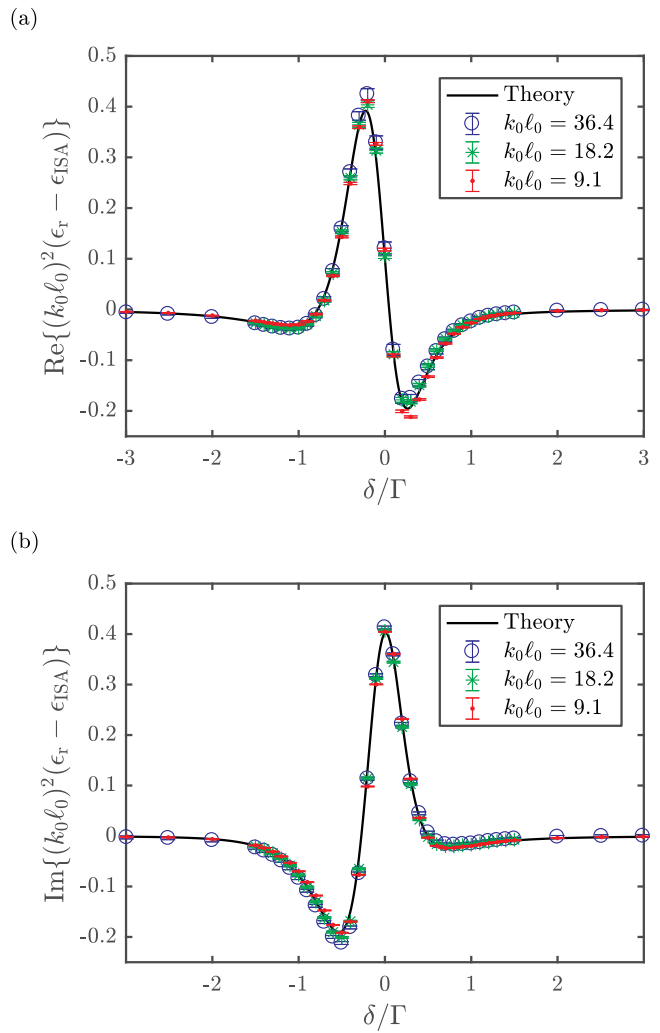}
   \caption{\label{fig:scalar2}   (a) Real part and (b) imaginary part of
   $(k_0\ell_0)^2\left(\epsilon_\text{r}-\epsilon_{\text{\tiny{ISA}}}\right)$ for scalar waves. The theoretical
   curve is shown as the black solid line. The numerical results at three different $k_0\ell_0$ values agree perfectly with the theoretical prediction.}
\end{figure}

In order to better compare our numerical results to the theoretical prediction, we calculate the scaled second order
contribution $(k_0\ell_0)^2 (\epsilon_\text{r} - \epsilon_{\text{\tiny{ISA}}})$. This quantity is calculated from the
numerical results, and compared to the theoretical prediction in~\fig{scalar2}. We find an excellent agreement between the numerical and theoretical results. The $(k_0\ell_0)^2
(\epsilon_\text{r} - \epsilon_{\text{\tiny{ISA}}})$ values at different $k_0\ell_0$ fall on the same curve, meaning
that the dominant contribution after ISA indeed scales as $1/k_0^2\ell_0^2$.

\subsection{Vector waves}
% -----------------------

\begin{figure}[h]
   \centering
   \includegraphics[width = \linewidth]{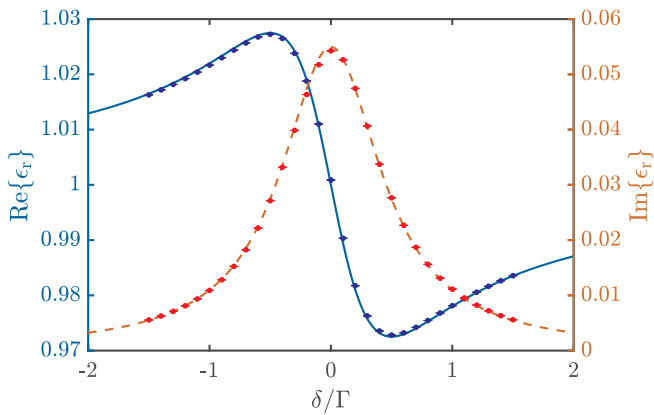}
   \caption{\label{fig:vector1} Real part and imaginary
   part of $\epsilon_\text{r}$ for the vector waves calculated from the numerical simulations at $k_0\ell_0=18.2$. The
   blue dots are the numerical results for the real part, while the red dots are the imaginary part. The ISA
   prediction for the real part is shown as the blue solid curve. The ISA prediction for the imaginary part is shown as
   the red dashed curve. The agreement is very good, but small deviations are visible and studied in \fig{vector2}.}
\end{figure}

For the vector waves, we set $2a =
\unit{22.5}{\micro\meter}$ ($2k_0a =409.5$), $2R = \unit{35}{\micro\meter}$ ($2k_0R = 637$), $w_0 =
\unit{4.5}{\micro\meter}$ ($k_0w_0=41$), $N_\text{conf} = 3200$, $N_\text{part} = 40$, $\tilde{d}_\text{min} = 0.455$
and the cylinder thickness is $L$=1, 2 and 4{\micro\meter}.   The corresponding scaled density of the cylinder is given
in Table~\ref{tab.scalarvector}, for the three values of $k_0\ell_0$. Generally, the numerical simulation is performed at $b_0=1$ or equivalently $k_0L=k_0\ell_0$,
with $N_a=2897$. A case at $b_0=2$ is also studied, where $k_0L=2k_0\ell_0$ and $N_a=5795$.  In~\fig{vector1}, we
compare the relative permittivity obtained by the numerical study at $k_0\ell_0=18.2$ to the theoretical prediction
using ISA, finding small but significant differences.

%
%\begingroup
%   \begin{table}
%      \caption{\label{tab.scalarvector} Values of $k_0\ell_0$ and corresponding $\rho_0/k_0^3$ values for the cases studied
%      numerically for the vector waves. The optical thickness at resonance is $b_0=1$ so that the cylinder
%      thickness is $L=\ell_0.$}
%      \begin{tabular}{c c c c c c}
%      \hline\hline
%         $k_0\ell_0$ & $\rho_0/k_0^3$\\
%      \hline
%         9.1 & $5.8\times10^{-3}$\\
%         18.2 &  $2.9\times10^{-3}$\\
%         36.4 &  $1.5\times10^{-3}$\\
%      \hline\hline
%      \end{tabular}
%   \end{table}
%\endgroup

\subsubsection{$1/k_0^2\ell_0^2$ dependence}
% ........................................

\begin{figure}[h]
   \centering
   \includegraphics[width = \linewidth]{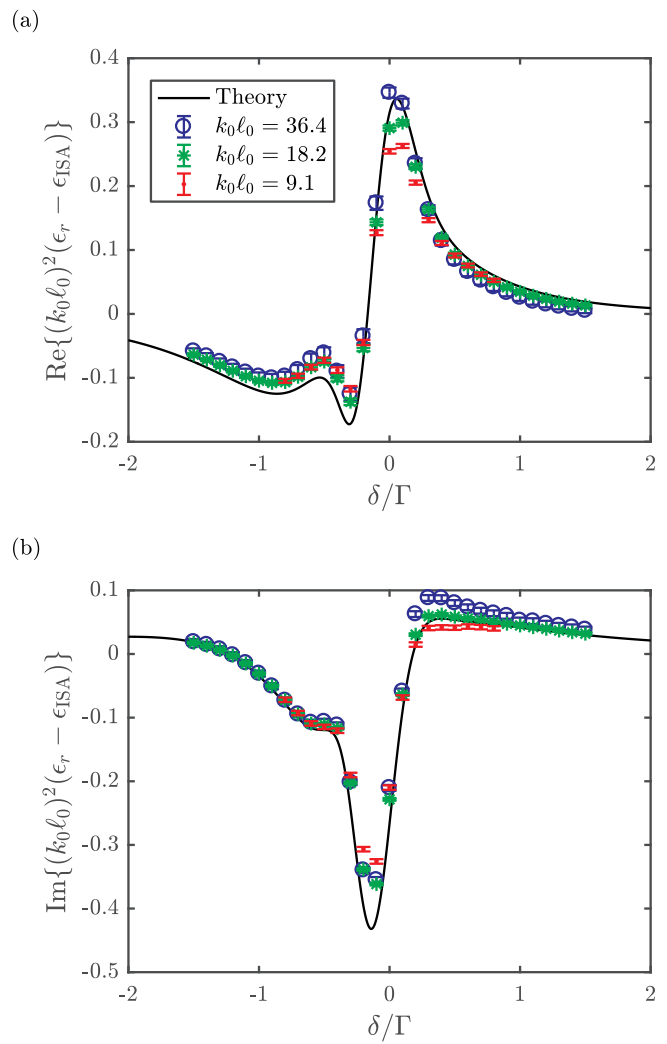}
   \caption{\label{fig:vector2}  (a) Real part and (b) imaginary part of
   $(k_0\ell_0)^2\left(\epsilon_\text{r}- \epsilon_{\text{\tiny{ISA}}}\right)$ for the case of vector waves. The
   theoretical curve is shown as the black solid curve. Numerical results are shown for three different $k_0\ell_0$
   values indicated in the legend.}
\end{figure}

We also compare the scaled second order contribution $(k_0\ell_0)^2\left(\epsilon_\text{r} -
\epsilon_{\text{\tiny{ISA}}}\right)$ of the numerical result to the theoretical prediction. This is shown
in~\fig{vector2}, where the numerically calculated second order contribution at $k_0\ell_0 = 9.1$, $18.2$ and $36.4$,
are scaled by multiplication with $(k_0\ell_0)^2$. Note that the numerical results for different values of $k_0\ell_0$
agree very well with each other, confirming the $1/k_0^2\ell_0^2$ dependence of $\epsilon_\text{r} -
\epsilon_{\text{\tiny{ISA}}}$. The agreement with the theoretical curve is not perfect, with small
deviations visible in the real part at negative detuning and in the imaginary part at small positive $\delta/\Gamma.$
Overall, the agreement is very good and the theoretical prediction nicely reproduces the complicated frequency
dependence, validating the theoretical approach.

The discrepancy between the numerical result and theoretical prediction is larger for $k_0\ell_0=18.2$ at certain values
of $\delta$ especially around $\delta/\Gamma=0.25$. It seems unlikely to come from correction terms proportional to
$1/k_0^3\ell_0^3$. The actual cause of this discrepancy remains to be understood. One reason could be the failure of the
bulk approximation in our numerical studies, since we are using the bulk permittivity for a medium where the thickness
is not much larger than the wavelength of the light. In the scalar case, the asymptotic expression of the Green function
is in fact valid at any distance [see \eq{G0}]. In the vector case, we speculate that corrections at short distance [see
\eq{vecGfree}] might lead to a less accurate bulk approximation. 

\subsubsection{Optical thickness}
% ................................

\begin{figure}[h]
   \centering
   \includegraphics[width = \linewidth]{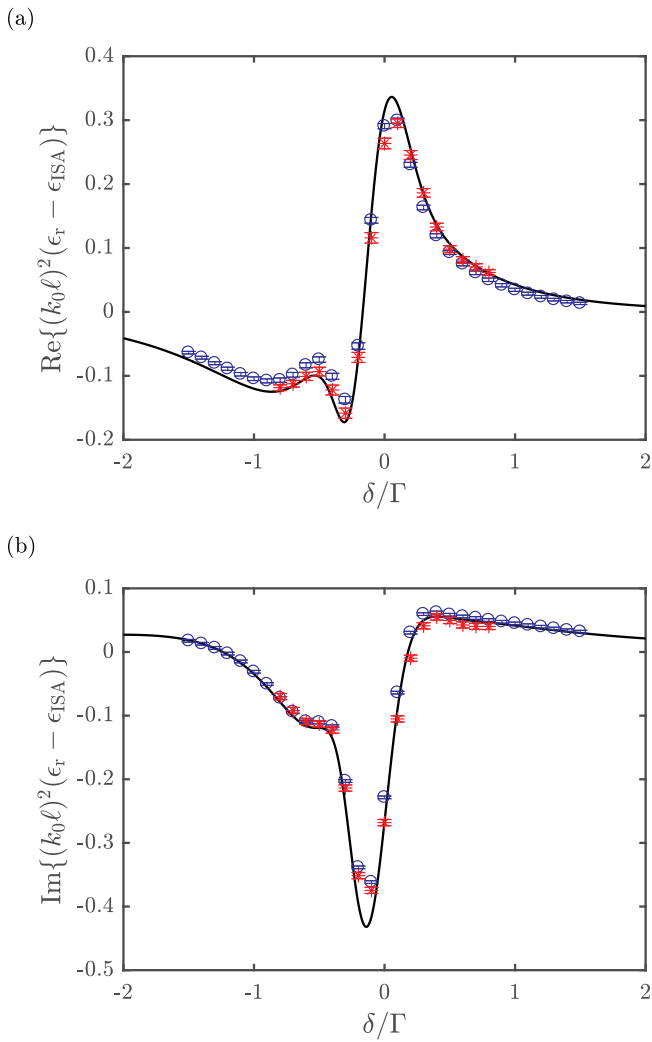}
   \caption{\label{fig:vectorb}  Numerical results for vector waves and for the case of $k_0\ell_0 = 18.2$ at optical thickness
   $b_0=1$ (blue open circles) and $b_0=2$ (red stars). The theoretical prediction of
   $(k_0\ell_0)^2\left(\epsilon_\text{r} - \epsilon_{\text{\tiny{ISA}}}\right)$ is shown as the black solid curve.}
\end{figure}

To check for possible finite size effect, we study the dependence of the second order contribution with
the thickness $L$. To do this, we compare the relative permittivity at the same value of $k_0\ell_0
=18.2$ for two different values of the optical thickness $b_0=L/\ell_0$, that is $b_0=1$ and $b_0=2$. The results are depicted in~\fig{vectorb}, showing good agreement between the two cases. Hence, the thickness used in the numerical simulation is sufficiently large and
finite size effects are not important. The excellent agreement between theory and numerical studies in the case of
scalar waves, where the geometry of the cylinder is similar, adds further weight to this conclusion.

\subsubsection{Size of the exclusion volume}
% ..........................................

\begin{figure}[h]
   \centering
   \includegraphics[width = \linewidth]{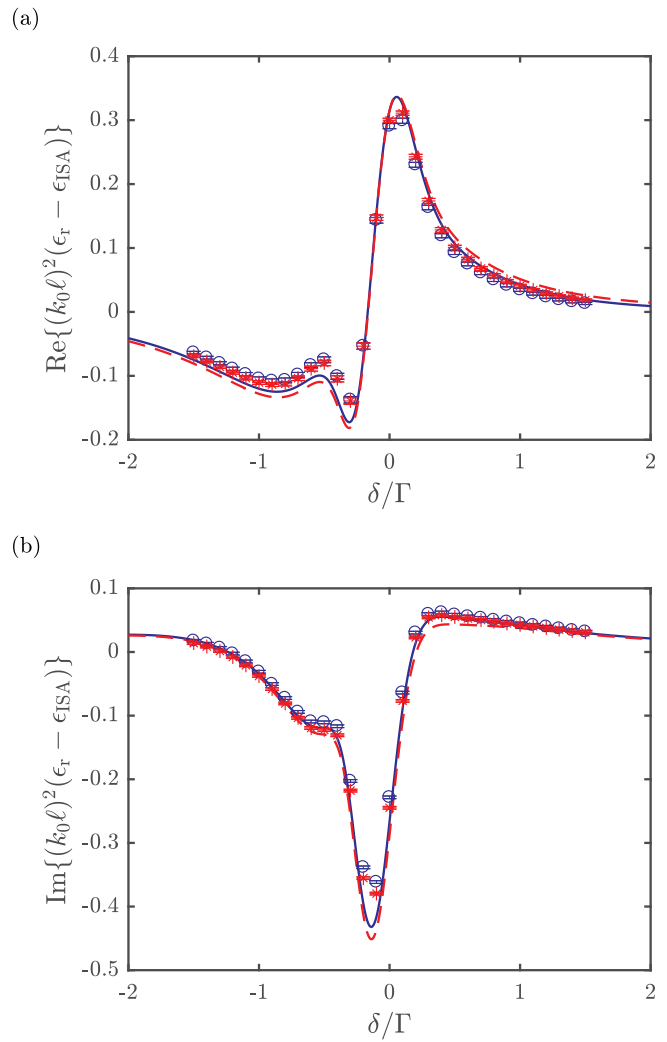}
   \caption{\label{fig:vectorc}  Numerical results for vector waves at $k_0\ell_0 = 18.2$, with normalized
   cut-off radii of $\tilde{d}_{\text{min}} = 0.455$ (blue open circles) and $\tilde{d}_{\text{min}} = 0.0455$ (red
   stars). The theoretical curves for $\tilde{d}_{\text{min}} = 0.455$ is shown as the black solid curve and the curve for
   $\tilde{d}_{\text{min}} = 0.0455$ is shown as the red dashed curve.}
\end{figure}

We also investigate the effect of the size of the exclusion volume in the case $k_0\ell_0 = 18.2$. We numerically study
the case of $\tilde{d}_\text{min} = 0.0455$, where the radius of exclusion volume is one order of magnitude smaller than
the results presented in~\fig{vector2}. The exclusion volume is thus three orders of magnitude smaller.
In~\fig{vectorc}, the values of $\epsilon_\text{r}$ for $\tilde{d}_\text{min} = 0.455$ and $\tilde{d}_\text{min} =
0.0455$ are compared. The results show that the smaller exclusion volume does not significantly affect the numerical and
the theoretical results.

\section{Conclusion}
% ==================

We have computed the relative permittivity of a bulk atomic cloud at zero temperature under illumination by a weak
probe beam, beyond the ISA. Using a diagrammatic representation method, we have identified all diagrams that contribute to the
self-energy up to second order in $1/k_0\ell_0$. The first order term in the self-energy gives rise to ISA, which is a
good approximation for dilute scattering media. The second order terms originate from the pairwise correlation in the
position of the scatterers, and from the recurrent scattering between two scatterers, which includes the well-known
Lorentz-Lorenz shift.   We have separately computed the contributions from ISA, position correlation and recurrent scattering for both the cases of scalar  and vector waves.

This perturbative expansion method can be useful to study non-trivial pair correlations. In Ref.~\onlinecite{bons2016}, the experimental measurement of the refractive index for a Bose gas,  was compared with two different models of bosonic pair correlation functions~\cite{naraschewski1999}, one for an ideal Bose gas, and the other one calculated with the Hartree-Fock approximation. A similar study could be carried out for Fermi gases.

From our calculations, we find that the peak optical thickness is shifted to the blue by $1.66\rho k_0^{-3}\Gamma$ and $6.56\rho k_0^{-3}\Gamma$, for the scalar wave and vector wave cases, respectively. A blue shift was also reported in Ref.~\onlinecite{corman2017}, for the case of quasi-2D atomic layers. 

The theoretical results are then compared with numerical studies on
finite-sized systems that approximate as closely as possible the infinite ideal slab geometry, with $k_0\ell_0$ values down to 9.1. The agreement between the
numerical and theoretical results is  almost perfect for the scalar waves. In the case of vector
waves, the overall agreement between the numerical and theoretical results is very good, although
differences exist. Further work is needed to understand these differences. Finally, our theoretical framework and numerical tools can be easily extended to study the effect of
the atomic motion on the coherent transmission of light beyond the ISA.

\begin{acknowledgments}
   CCK is grateful to CQT and ESPCI for funding his trip to Paris. RP acknowledges the support of LABEX WIFI (Laboratory
   of Excellence ANR-10-LABX-24) within the French Program ``Investments for the Future'' under reference
   ANR-10-IDEX-0001-02 PSL$^{\ast}$. DD thanks Nicolas Cherroret for enlightning discussions. CCK thanks Janne Ruostekoski for   
   helpful comments. This work was supported by the CQT/MoE funding Grant No. R-710-002-016-271.
\end{acknowledgments}

% =======
\appendix
% =======

\section{Calculation of $\mscor$ for scalar waves}\label{app.scalar1}
% ================================================

Starting from~\eq{scorr}, we perform the integrations and summations to find
\begin{equation}
   \mscor(\bm{r}-\bm{r}') = \frac{N_a(N_a-1)}{V^2}G_0(\bm{r}-\bm{r}') h(|\bm{r}-\bm{r}'|).
\end{equation}
Since $N_a\gg1$, we can make the approximation that $N_a(N_a-1) \approx N_a^2$. The
self-energy in $\bm{k}$-space is given by the following Fourier transform,
\begin{equation}
   \mscor(\bm{k}) = \rho^2\alpha^2k_L^4 \int G_0(\bm{R}') h(R')
   \exp(-i\bm{k}\cdot\bm{R}')\mathrm{d}^3\bm{R}',
\end{equation}
where $\bm{R}'=\bm{r}-\bm{r}'$. Putting in the pair correlation function from~\eq{paircorrfunc}, it is reduced to the following
integral over the exclusion volume $V_{\text{ex}}$,
\begin{equation}
   \mscor(\bm{k}) = -\rho^2\alpha^2k_L^4\int_{V_\text{ex}} G_0(\bm{R}')\exp(-i\bm{k}\cdot\bm{R}')\mathrm{d}^3\bm{R}'.\label{eq:scie}
\end{equation}
The scalar Green function does not contain any angular dependance, therefore it is possible to first carry out
the integral over the solid angle, leading to
\begin{equation}
   \mscor(\bm{k}) = -\rho^2\alpha^2k_L^4\int_0^{d_\text{min}}\frac{\exp(-ik_LR')}{R'}\frac{\sin kR'}{kR'} R'^2 \mathrm{d}R'.
\end{equation}
We now make the on-shell approximation to put $k\approx k_L$. We then perform the integration to arrive at~\eq{scorrfinal}.

\section{Calculation of $\msrec$ for scalar waves}\label{app.scalar2}
% ================================================

In close analogy to~\eqs{srec3}{srec4}, we can write $\msrec^{(n)}$ for general values of $n$ in the configuration
space. In the Fourier space, this is given by
\begin{multline}
   \msrec^{(n)}(\bm{k}) = \rho^2 (\alpha k_L^2)^n\\
   \times \begin{cases}
      \int G_0^{n-1}(\bm{R}')[1+h(R')]\mathrm{d}^3\bm{R}', & \text{odd $n$},\\
      \int G_0^{n-1}(\bm{R}')[1+h(R')]\exp(-i\bm{k}\cdot\bm{R}')\mathrm{d}^3\bm{R}', & \text{even $n$},\\
   \end{cases}
\end{multline}
for $n\ge3$. We have used the fact that $G_0(\bm{R}') = G_0(-\bm{R}')$, which is also true for the dyadic
Green function in the vector case. Note also that $N_a(N_a-1)\approx N_a^2$ for large $N_a$. Summing up
$\msrec^{(n)}$ and putting in the pair correlation function of ~\eq{paircorrfunc}, we obtain~\eq{totalrec}.

To proceed from~\eq{totalrec}, we first carry out the angular integration, with the on-shell approximation, to obtain the following equation:
\begin{multline}
   \msrec(\bm{k})= 4\pi(\rho\alpha_0)^2 k_L^4 \sum_{l=0}^{\infty} \left(\alpha_0 k_L^2\right)^{2l+1}\tilde{\alpha}^{2l+3}\\
   \times\Bigg[\int_{d_\text{min}}^\infty G_0^{2l+2}(\bm{R}')R'^2\mathrm{d}R'+\left(\alpha_0 k_L^2\right)\tilde{\alpha}\\
   \times\int_{d_\text{min}}^\infty G_0^{2l+3}(\bm{R}')\frac{\sin kR'}{kR'}R'^2\mathrm{d}R'\Bigg].
\end{multline}
 We also make the approximation $k_L \approx k_0$,
for $\delta \ll \omega_0$. Thus, $\alpha_0k_L^3 = 4\pi$, allowing us to simplify the equation, and find
\begin{multline}
   \frac{\msrec}{k_L^2} = \frac{1}{(k_0\ell_0)^2}\left[\sum_{l=0}^{\infty}\tilde{\alpha}^{2l+3}
      \int_{\tilde{d}_\text{min}}^\infty\frac{e^{i(2l+2)x}}{x^{2l}}\mathrm{d}x\right.\\
   \left.+ \sum_{l=0}^{\infty}\tilde{\alpha}^{2l+4}\int_{\tilde{d}_\text{min}}^\infty\frac{e^{i(2l+3)x}}{x^{2l+2}}\sin x\mathrm{d}x\right].
\end{multline}
Here, we have a summation of infinitely many integrals. The first integral proportional to $\tilde{\alpha}^3$ can be computed analytically,
\begin{equation}
   \int_{\tilde{d}_\text{min}}^\infty \exp(2ix)\mathrm{d}x = \lim_{\eta\rightarrow0^+}\int_{\tilde{d}_\text{min}}^\infty\exp[(2i-\eta)x]\mathrm{d}x,
\end{equation}
which evaluates to the value $i\exp(2i\tilde{d}_\text{min})/2$ for ${\eta >0}$. After reorganizing the summations, we have
\begin{multline}
   \frac{\msrec}{k_L^2} = \frac{1}{(k_0\ell_0)^2}\frac{i\tilde{\alpha}^3 }{2} \Bigg[e^{2i\tilde{d}_\text{min}} \\
   +\sum_{l=0}^{\infty}\tilde{\alpha}^{2l+1}\int_{\tilde{d}_\text{min}}^\infty \frac{e^{(2l+2)ix}}{x^{2l+2}}\mathrm{d}x\\
   -(1+2i\tilde{\alpha})\sum_{l=0}^{\infty}\tilde{\alpha}^{2l+1} \int_{\tilde{d}_\text{min}}^\infty \frac{e^{(2l+4)ix}}{x^{2l+2}}\mathrm{d}x\Bigg].
\end{multline}
Finally, to obtain~\eq{recscalar}, we interchange the order of the summation and the integration in the above equation.

\section{Calculation of $\mvcor$ for vector waves}\label{app.vector1}
% ================================================

As discussed in the text, we have to use the dyadic version of~\eq{scie} to compute $\mvcor$, which is,
\begin{equation}
   \mvcor(\bm{k}) = -\rho^2\alpha^2k_L^4\int_{V_\text{ex}} \tensor{\bm{G}}_0(\bm{R}')\exp(-i\bm{k}\cdot\bm{R}')\mathrm{d}^3\bm{R}'.
\end{equation}
Putting in the dyadic Green function for vector waves,
\begin{multline}
   \mvcor(\bm{k}) = -\rho^2\alpha^2k_L^4\Bigg[\frac{k_L}{4\pi}\int_{V_\text{ex}} \beta(k_LR')
      \tensor{\bm{P}} e^{-i\bm{k}\cdot\bm{R}'} \mathrm{d}^3\bm{R}'\\
   + \frac{k_L}{4\pi}\int_{V_\text{ex}} \gamma(k_LR')\tensor{\bm{U}} e^{-i\bm{k}\cdot\bm{R}'} \mathrm{d}^3\bm{R}'\\
   - \int_{V_\text{ex}} \frac{\tensor{\bm{I}}}{3k_L^2}\delta(\bm{R}') e^{-i\bm{k}\cdot\bm{R}'}
      \mathrm{d}^3\bm{R}'\Bigg]\label{eq:m2c}.
\end{multline}
The last term can be computed easily using the properties of the Dirac delta function,
\begin{equation}
   \int_{V_\text{ex}} \frac{\tensor{\bm{I}}}{3k_L^2}\delta(\bm{R}') e^{-i\bm{k}\cdot\bm{R}'} \mathrm{d}^3\bm{R}'
      = \frac{\tensor{\bm{I}}}{3k_L^2}.
\end{equation}
Before we proceed to evaluate the two integrals involving $\beta(k_LR')$ and $\gamma(k_LR')$, we note the following
results for the integration of $\tensor{\bm{P}}$ and $\tensor{\bm{U}}$ over the solid angle $\Omega$. These
relations are useful in the computation of $\mvcor$ and $\mvrec$.
\begin{align}
   \int \tensor{\bm{U}} \mathrm{d}\Omega =& \frac{4\pi}{3}\tensor{\bm{I}},\label{eq:iU}\\
   \int \tensor{\bm{P}} \mathrm{d}\Omega =& \frac{8\pi}{3}\tensor{\bm{I}},\label{eq:iP}\\
   \int \tensor{\bm{U}} e^{-i \bm{k}\cdot \bm{R}} \mathrm{d}\Omega =& 4\pi\frac{j_1\left(k R\right)}{k R} \tensor{\bm{I}}\nonumber\\
      & + \left[4\pi j_0\left(kR\right)-12\pi\frac{j_1\left(k R\right)}{k R}\right]\tensor{\bm{K}},\label{eq:iUeikR}\\
   \int \tensor{\bm{P}} {e}^{-i \bm{k}\cdot \bm{R}} \mathrm{d}\Omega =&
      \left[4\pi j_0(kR)-4\pi\frac{j_1\left(k R\right)}{k R}\right] \tensor{\bm{I}}\nonumber\\
   & -\left[4\pi j_0\left(kR\right)-12\pi\frac{j_1\left(k R\right)}{k R}\right]\tensor{\bm{K}}.\label{eq:iPeikR}
\end{align}
Using these, we can perform the angular integration over the solid angle of~\eq{m2c}. The following transverse component
of the mass operator is what matters to us here,
\begin{align}
   \mtcor(\bm{k}) =& \frac{\rho^2 \alpha^2k_L^2}{3}-\rho^2\alpha^2k_L^5 \nonumber\\
   \times& \Bigg[\int_0^{d_\text{min}} \!\!\gamma(k_LR') \frac{j_1(kR')}{kR'} R'^2\,\mathrm{d} R'\nonumber\\
   +&\int_0^{d_\text{min}}\!\! \beta(k_LR') \frac{kR'j_0(kR')-j_1(kR')}{kR'} R'^2\,\mathrm{d} R'\Bigg].
\end{align}
Applying the on-shell approximation, and evaluating the remaining integral over $R'$ leads us to~\eqs{posc}{posc2}.

\section{Calculation of $\mvrec$ for vector waves}\label{app.vector2}
% ================================================

\subsection{Derivation}
% ---------------------

In order to calculate $\mvrec$, we consider the following equation which is the dyadic version of~\eq{totalrec},
\begin{multline}
   \mvrec(\bm{k})= (\rho\alpha_0)^2 k_L^4 \sum_{l=0}^{\infty} \left(\alpha_0 k_L^2\right)^{2l+1}\tilde{\alpha}^{2l+3}\\
   \times\Bigg[\int_{V-V_\text{ex}} \tensor{\bm{G}}_0^{2l+2}(\bm{R}')\mathrm{d}^3\bm{R}'+\left(\alpha_0 k_L^2\right)\tilde{\alpha}\\
   \times\int_{V-V_\text{ex}} \tensor{\bm{G}}_0^{2l+3}(\bm{R}')e^{-i\bm{k}\cdot\bm{R}'}\mathrm{d}^3\bm{R}'\Bigg].\label{eq:m2r}
\end{multline}
$\tensor{\bm{U}}$ and $\tensor{\bm{P}}$ being orthogonal projectors, one has
$\tensor{\bm{U}}^2 = \tensor{\bm{U}}$, $\tensor{\bm{P}}^2 = \tensor{\bm{P}}$ and $\tensor{\bm{U}}\tensor{\bm{P}} =
\tensor{\bm{P}}\tensor{\bm{U}} = \tensor{0}$. It then follows that the $n$-th power of the vector Green function is
given by
\begin{equation}
   \tensor{\bm{G}}_0^n(\bm{R}') = \left(\frac{k_L}{4\pi}\right)^{\!\!n}\left(\beta^n(k_LR')\tensor{\bm{P}} + \gamma^n(k_LR')\tensor{\bm{U}}\right).
\end{equation}
We have neglected the Dirac delta term in $\tensor{\bm{G}}$, since the integration over the volume $V-V_{\text{ex}}$
excludes the origin. The above relation is substituted into~\eq{m2r}. An integration over the solid angle is first performed, making use of~\eqr{iU}{iPeikR}. The transverse component of the result is given by
\begin{align}
   &\mtrec(\bm{k}) = \rho^2\alpha_0^2k_L^5\sum_{l=0}^\infty \left(\frac{\alpha_0 k_L^3}{4\pi}\right)^{2l+1}\!\tilde{\alpha}^{2l+3}\nonumber\\
   &\times \Bigg[\frac{1}{3}\int_{d_\text{min}}^\infty \left[2\beta^{2l+2}(k_LR')+\gamma^{2l+2}(k_LR')\right] R'^2 \,\mathrm{d} R' \nonumber\\
   &+ \frac{\alpha_0k_L^3}{4\pi}\tilde{\alpha}\int_{d_\text{min}}^\infty \beta^{2l+3}(k_L R')\frac{kR'j_0(kR')-j_1(kR')}{kR'} R'^2 \,\mathrm{d} R'\nonumber\\
   &+ \frac{\alpha_0k_L^3}{4\pi}\tilde{\alpha}\int_{d_\text{min}}^\infty\gamma^{2l+3}(k_L R')\frac{j_1(kR')}{kR'} R'^2 \,\mathrm{d} R'\Bigg].
\end{align}

We then assume that $k\approx k_L$ and define $x = k_LR'$. For vector waves,  $\alpha_0k_L^3=6\pi$, assuming $k_0\approx k_L$.  This leads us to
\begin{align}
\mtrec = &\rho^2\alpha_0^2 k_L^2\sum_{l=0}^\infty{\left(\frac{3}{2}\right)}^{2l+1}\tilde{\alpha}^{2l+3}\nonumber\\
 &\times\Bigg\{\frac{1}{3}\int_{\tilde{d}_\text{min}}^\infty\left[2\beta^{2l+2}(x)+\gamma^{2l+2}(x)\right]x^2\,\mathrm{d} x\nonumber\\
 &+ \frac{3}{2}\tilde{\alpha} \int_{\tilde{d}_\text{min}}^\infty \beta^{2l+3}(x)x^2\left[j_0(x)-j_1(x)/x\right]\mathrm{d} x\nonumber\\
 &+ \frac{3}{2} \tilde{\alpha} \int_{\tilde{d}_\text{min}}^\infty \gamma^{2l+3}(x)xj_1(x)\,\mathrm{d} x\Bigg\}.
\end{align}

Next, we interchange the order of integration and summation, to arrive at~\eqs{vrec1}{vrec2}.

\subsection{Numerical integration}
% --------------------------------

A numerical integration is performed to compute $I_v$. The same numerical algorithm used for the scalar case is also used for the vector waves. First, $I_v$ is separated into two parts
\begin{equation}
I_v = I_v' + I_v^M,
\end{equation}
with an large value of $M$. One part is given by the following integral
\begin{multline}
   I_v^M =  \int_{\tilde{d}_\text{min}}^M \!\frac{2x^2\left[\beta(x)^2 + \frac{9}{4}\tilde{\alpha}(\omega)\beta(x)^3\left\{j_0(x)-j_1(x)/x\right\}\right]}{1-\frac{9}{4}\tilde{\alpha}(\omega)^2 \beta(x)^2}\\+x^2\frac{\gamma(x)^2 + \frac{9}{2}\tilde{\alpha}(\omega)\gamma(x)^3j_1(x)/x}{1-\frac{9}{4}\tilde{\alpha}(\omega)^2 \gamma(x)^2}\,\mathrm{d} x. \label{eq:vrec3}
\end{multline}
which is calculated numerical using the adaptive algorithm. The value of $M$ is set at 10000. The remaining part $I_v'$ can be computed analytically for large value of $M$, giving
\begin{equation}
I_v' \approx \lim_{\eta\rightarrow0^+}\int_M^\infty 2\mathrm{e}^{2ix-\eta x}\,\mathrm{d} x = i \exp(2iM).
\end{equation}

The final result is of course almost independent of the intermediate $M$ value.

% Biblio
\newcommand{\noopsort}[1]{}\providecommand{\noopsort}[1]{}\providecommand{\singleletter}[1]{#1}%


\begin{thebibliography}{52}%
\makeatletter
\providecommand \@ifxundefined [1]{%
 \@ifx{#1\undefined}
}%
\providecommand \@ifnum [1]{%
 \ifnum #1\expandafter \@firstoftwo
 \else \expandafter \@secondoftwo
 \fi
}%
\providecommand \@ifx [1]{%
 \ifx #1\expandafter \@firstoftwo
 \else \expandafter \@secondoftwo
 \fi
}%
\providecommand \natexlab [1]{#1}%
\providecommand \enquote  [1]{``#1''}%
\providecommand \bibnamefont  [1]{#1}%
\providecommand \bibfnamefont [1]{#1}%
\providecommand \citenamefont [1]{#1}%
\providecommand \href@noop [0]{\@secondoftwo}%
\providecommand \href [0]{\begingroup \@sanitize@url \@href}%
\providecommand \@href[1]{\@@startlink{#1}\@@href}%
\providecommand \@@href[1]{\endgroup#1\@@endlink}%
\providecommand \@sanitize@url [0]{\catcode `\\12\catcode `\$12\catcode
  `\&12\catcode `\#12\catcode `\^12\catcode `\_12\catcode `\%12\relax}%
\providecommand \@@startlink[1]{}%
\providecommand \@@endlink[0]{}%
\providecommand \url  [0]{\begingroup\@sanitize@url \@url }%
\providecommand \@url [1]{\endgroup\@href {#1}{\urlprefix }}%
\providecommand \urlprefix  [0]{URL }%
\providecommand \Eprint [0]{\href }%
\providecommand \doibase [0]{http://dx.doi.org/}%
\providecommand \selectlanguage [0]{\@gobble}%
\providecommand \bibinfo  [0]{\@secondoftwo}%
\providecommand \bibfield  [0]{\@secondoftwo}%
\providecommand \translation [1]{[#1]}%
\providecommand \BibitemOpen [0]{}%
\providecommand \bibitemStop [0]{}%
\providecommand \bibitemNoStop [0]{.\EOS\space}%
\providecommand \EOS [0]{\spacefactor3000\relax}%
\providecommand \BibitemShut  [1]{\csname bibitem#1\endcsname}%
\let\auto@bib@innerbib\@empty
%</preamble>
\bibitem [{\citenamefont {Fioretti}\ \emph {et~al.}(1998)\citenamefont
  {Fioretti}, \citenamefont {Molisch}, \citenamefont {M{\"u}ller},
  \citenamefont {Verkerk},\ and\ \citenamefont {Allegrini}}]{fioretti1998}%
  \BibitemOpen
  \bibfield  {author} {\bibinfo {author} {\bibfnamefont {A.}~\bibnamefont
  {Fioretti}}, \bibinfo {author} {\bibfnamefont {A.~F.}\ \bibnamefont
  {Molisch}}, \bibinfo {author} {\bibfnamefont {J.~H.}\ \bibnamefont
  {M{\"u}ller}}, \bibinfo {author} {\bibfnamefont {P.}~\bibnamefont {Verkerk}},
  \ and\ \bibinfo {author} {\bibfnamefont {M.}~\bibnamefont {Allegrini}},\
  }\href@noop {} {\bibfield  {journal} {\bibinfo  {journal} {Opt. Commun.}\
  }\textbf {\bibinfo {volume} {149}},\ \bibinfo {pages} {415} (\bibinfo {year}
  {1998})}\BibitemShut {NoStop}%
\bibitem [{\citenamefont {Labeyrie}\ \emph {et~al.}(2003)\citenamefont
  {Labeyrie}, \citenamefont {Vaujour}, \citenamefont {M{\"u}ller},
  \citenamefont {Delande}, \citenamefont {Miniatura}, \citenamefont
  {Wilkowski},\ and\ \citenamefont {Kaiser}}]{labeyrie2003}%
  \BibitemOpen
  \bibfield  {author} {\bibinfo {author} {\bibfnamefont {G.}~\bibnamefont
  {Labeyrie}}, \bibinfo {author} {\bibfnamefont {E.}~\bibnamefont {Vaujour}},
  \bibinfo {author} {\bibfnamefont {C.~A.}\ \bibnamefont {M{\"u}ller}},
  \bibinfo {author} {\bibfnamefont {D.}~\bibnamefont {Delande}}, \bibinfo
  {author} {\bibfnamefont {C.}~\bibnamefont {Miniatura}}, \bibinfo {author}
  {\bibfnamefont {D.}~\bibnamefont {Wilkowski}}, \ and\ \bibinfo {author}
  {\bibfnamefont {R.}~\bibnamefont {Kaiser}},\ }\href@noop {} {\bibfield
  {journal} {\bibinfo  {journal} {Phys. Rev. Lett.}\ }\textbf {\bibinfo
  {volume} {91}},\ \bibinfo {pages} {223904} (\bibinfo {year}
  {2003})}\BibitemShut {NoStop}%
\bibitem [{\citenamefont {Labeyrie}\ \emph {et~al.}(1999)\citenamefont
  {Labeyrie}, \citenamefont {de~Tomasi}, \citenamefont {Bernard}, \citenamefont
  {M{\"u}ller}, \citenamefont {Miniatura},\ and\ \citenamefont
  {Kaiser}}]{labeyrie1999}%
  \BibitemOpen
  \bibfield  {author} {\bibinfo {author} {\bibfnamefont {G.}~\bibnamefont
  {Labeyrie}}, \bibinfo {author} {\bibfnamefont {F.}~\bibnamefont {de~Tomasi}},
  \bibinfo {author} {\bibfnamefont {J.-C.}\ \bibnamefont {Bernard}}, \bibinfo
  {author} {\bibfnamefont {C.~A.}\ \bibnamefont {M{\"u}ller}}, \bibinfo
  {author} {\bibfnamefont {C.}~\bibnamefont {Miniatura}}, \ and\ \bibinfo
  {author} {\bibfnamefont {R.}~\bibnamefont {Kaiser}},\ }\href@noop {}
  {\bibfield  {journal} {\bibinfo  {journal} {Phys. Rev. Lett.}\ }\textbf
  {\bibinfo {volume} {83}},\ \bibinfo {pages} {5266} (\bibinfo {year}
  {1999})}\BibitemShut {NoStop}%
\bibitem [{\citenamefont {Bidel}\ \emph {et~al.}(2002)\citenamefont {Bidel},
  \citenamefont {Klappnauf}, \citenamefont {Bernard}, \citenamefont {Delande},
  \citenamefont {Labeyrie}, \citenamefont {Miniatura}, \citenamefont
  {Wilkowski},\ and\ \citenamefont {Kaiser}}]{bidel2002}%
  \BibitemOpen
  \bibfield  {author} {\bibinfo {author} {\bibfnamefont {Y.}~\bibnamefont
  {Bidel}}, \bibinfo {author} {\bibfnamefont {B.}~\bibnamefont {Klappnauf}},
  \bibinfo {author} {\bibfnamefont {J.~C.}\ \bibnamefont {Bernard}}, \bibinfo
  {author} {\bibfnamefont {D.}~\bibnamefont {Delande}}, \bibinfo {author}
  {\bibfnamefont {G.}~\bibnamefont {Labeyrie}}, \bibinfo {author}
  {\bibfnamefont {C.}~\bibnamefont {Miniatura}}, \bibinfo {author}
  {\bibfnamefont {D.}~\bibnamefont {Wilkowski}}, \ and\ \bibinfo {author}
  {\bibfnamefont {R.}~\bibnamefont {Kaiser}},\ }\href@noop {} {\bibfield
  {journal} {\bibinfo  {journal} {Phys. Rev. Lett.}\ }\textbf {\bibinfo
  {volume} {88}},\ \bibinfo {pages} {203902} (\bibinfo {year}
  {2002})}\BibitemShut {NoStop}%
\bibitem [{\citenamefont {Baudouin}\ \emph {et~al.}(2013)\citenamefont
  {Baudouin}, \citenamefont {Mercadier}, \citenamefont {Guarrera},
  \citenamefont {Gu{\'e}rin},\ and\ \citenamefont {Kaiser}}]{baudouin2013}%
  \BibitemOpen
  \bibfield  {author} {\bibinfo {author} {\bibfnamefont {Q.}~\bibnamefont
  {Baudouin}}, \bibinfo {author} {\bibfnamefont {N.}~\bibnamefont {Mercadier}},
  \bibinfo {author} {\bibfnamefont {V.}~\bibnamefont {Guarrera}}, \bibinfo
  {author} {\bibfnamefont {W.}~\bibnamefont {Gu{\'e}rin}}, \ and\ \bibinfo
  {author} {\bibfnamefont {R.}~\bibnamefont {Kaiser}},\ }\href@noop {}
  {\bibfield  {journal} {\bibinfo  {journal} {Nat. Phys.}\ }\textbf {\bibinfo
  {volume} {9}},\ \bibinfo {pages} {357} (\bibinfo {year} {2013})}\BibitemShut
  {NoStop}%
\bibitem [{\citenamefont {Chalony}\ \emph {et~al.}(2011)\citenamefont
  {Chalony}, \citenamefont {Pierrat}, \citenamefont {Delande},\ and\
  \citenamefont {Wilkowski}}]{chalony2011}%
  \BibitemOpen
  \bibfield  {author} {\bibinfo {author} {\bibfnamefont {M.}~\bibnamefont
  {Chalony}}, \bibinfo {author} {\bibfnamefont {R.}~\bibnamefont {Pierrat}},
  \bibinfo {author} {\bibfnamefont {D.}~\bibnamefont {Delande}}, \ and\
  \bibinfo {author} {\bibfnamefont {D.}~\bibnamefont {Wilkowski}},\ }\href@noop
  {} {\bibfield  {journal} {\bibinfo  {journal} {{Phys. Rev. A}}\ }\textbf
  {\bibinfo {volume} {{84}}},\ \bibinfo {pages} {011401(R)} (\bibinfo {year}
  {{2011}})}\BibitemShut {NoStop}%
\bibitem [{\citenamefont {Kwong}\ \emph {et~al.}(2014)\citenamefont {Kwong},
  \citenamefont {Yang}, \citenamefont {Pramod}, \citenamefont {Pandey},
  \citenamefont {Delande}, \citenamefont {Pierrat},\ and\ \citenamefont
  {Wilkowski}}]{kwong2014cooperative}%
  \BibitemOpen
  \bibfield  {author} {\bibinfo {author} {\bibfnamefont {C.~C.}\ \bibnamefont
  {Kwong}}, \bibinfo {author} {\bibfnamefont {T.}~\bibnamefont {Yang}},
  \bibinfo {author} {\bibfnamefont {M.~S.}\ \bibnamefont {Pramod}}, \bibinfo
  {author} {\bibfnamefont {K.}~\bibnamefont {Pandey}}, \bibinfo {author}
  {\bibfnamefont {D.}~\bibnamefont {Delande}}, \bibinfo {author} {\bibfnamefont
  {R.}~\bibnamefont {Pierrat}}, \ and\ \bibinfo {author} {\bibfnamefont
  {D.}~\bibnamefont {Wilkowski}},\ }\href@noop {} {\bibfield  {journal}
  {\bibinfo  {journal} {Phys. Rev. Lett.}\ }\textbf {\bibinfo {volume} {113}},\
  \bibinfo {pages} {223601} (\bibinfo {year} {2014})}\BibitemShut {NoStop}%
\bibitem [{\citenamefont {Kwong}\ \emph {et~al.}(2015)\citenamefont {Kwong},
  \citenamefont {Yang}, \citenamefont {Delande}, \citenamefont {Pierrat},\ and\
  \citenamefont {Wilkowski}}]{kwong2015cooperative}%
  \BibitemOpen
  \bibfield  {author} {\bibinfo {author} {\bibfnamefont {C.~C.}\ \bibnamefont
  {Kwong}}, \bibinfo {author} {\bibfnamefont {T.}~\bibnamefont {Yang}},
  \bibinfo {author} {\bibfnamefont {D.}~\bibnamefont {Delande}}, \bibinfo
  {author} {\bibfnamefont {R.}~\bibnamefont {Pierrat}}, \ and\ \bibinfo
  {author} {\bibfnamefont {D.}~\bibnamefont {Wilkowski}},\ }\href@noop {}
  {\bibfield  {journal} {\bibinfo  {journal} {Phys. Rev. Lett.}\ }\textbf
  {\bibinfo {volume} {115}},\ \bibinfo {pages} {223601} (\bibinfo {year}
  {2015})}\BibitemShut {NoStop}%
\bibitem [{\citenamefont {Bettles}\ \emph {et~al.}(2018)\citenamefont
  {Bettles}, \citenamefont {Ilieva}, \citenamefont {Busche}, \citenamefont
  {Huillery}, \citenamefont {Ball}, \citenamefont {Spong},\ and\ \citenamefont
  {Adams}}]{bettles2018faster}%
  \BibitemOpen
  \bibfield  {author} {\bibinfo {author} {\bibfnamefont {R.~J.}\ \bibnamefont
  {Bettles}}, \bibinfo {author} {\bibfnamefont {T.}~\bibnamefont {Ilieva}},
  \bibinfo {author} {\bibfnamefont {H.}~\bibnamefont {Busche}}, \bibinfo
  {author} {\bibfnamefont {P.}~\bibnamefont {Huillery}}, \bibinfo {author}
  {\bibfnamefont {S.~W.}\ \bibnamefont {Ball}}, \bibinfo {author}
  {\bibfnamefont {N.~L.}\ \bibnamefont {Spong}}, \ and\ \bibinfo {author}
  {\bibfnamefont {C.~S.}\ \bibnamefont {Adams}},\ }\href@noop {} {\bibfield
  {journal} {\bibinfo  {journal} {arXiv:1808.08415}\ } (\bibinfo {year}
  {2018})}\BibitemShut {NoStop}%
\bibitem [{\citenamefont {Balik}\ \emph {et~al.}(2013)\citenamefont {Balik},
  \citenamefont {Win}, \citenamefont {Havey}, \citenamefont {Sokolov},\ and\
  \citenamefont {Kupriyanov}}]{balik2013}%
  \BibitemOpen
  \bibfield  {author} {\bibinfo {author} {\bibfnamefont {S.}~\bibnamefont
  {Balik}}, \bibinfo {author} {\bibfnamefont {A.~L.}\ \bibnamefont {Win}},
  \bibinfo {author} {\bibfnamefont {M.~D.}\ \bibnamefont {Havey}}, \bibinfo
  {author} {\bibfnamefont {I.~M.}\ \bibnamefont {Sokolov}}, \ and\ \bibinfo
  {author} {\bibfnamefont {D.~V.}\ \bibnamefont {Kupriyanov}},\ }\href@noop {}
  {\bibfield  {journal} {\bibinfo  {journal} {Phys. Rev. A}\ }\textbf {\bibinfo
  {volume} {87}},\ \bibinfo {pages} {053817} (\bibinfo {year}
  {2013})}\BibitemShut {NoStop}%
\bibitem [{\citenamefont {Sokolov}(2015)}]{sokolov2015}%
  \BibitemOpen
  \bibfield  {author} {\bibinfo {author} {\bibfnamefont {I.~M.}\ \bibnamefont
  {Sokolov}},\ }\href {http://stacks.iop.org/1555-6611/25/i=6/a=065202}
  {\bibfield  {journal} {\bibinfo  {journal} {Laser Physics}\ }\textbf
  {\bibinfo {volume} {25}},\ \bibinfo {pages} {065202} (\bibinfo {year}
  {2015})}\BibitemShut {NoStop}%
\bibitem [{\citenamefont {Bienaim\'e}\ \emph {et~al.}(2012)\citenamefont
  {Bienaim\'e}, \citenamefont {Piovella},\ and\ \citenamefont
  {Kaiser}}]{bienaime2012}%
  \BibitemOpen
  \bibfield  {author} {\bibinfo {author} {\bibfnamefont {T.}~\bibnamefont
  {Bienaim\'e}}, \bibinfo {author} {\bibfnamefont {N.}~\bibnamefont
  {Piovella}}, \ and\ \bibinfo {author} {\bibfnamefont {R.}~\bibnamefont
  {Kaiser}},\ }\href {\doibase 10.1103/PhysRevLett.108.123602} {\bibfield
  {journal} {\bibinfo  {journal} {Phys. Rev. Lett.}\ }\textbf {\bibinfo
  {volume} {108}},\ \bibinfo {pages} {123602} (\bibinfo {year}
  {2012})}\BibitemShut {NoStop}%
\bibitem [{\citenamefont {Ara\'ujo}\ \emph {et~al.}(2016)\citenamefont
  {Ara\'ujo}, \citenamefont {Kre\ifmmode \check{s}\else
  \v{s}\fi{}i\ifmmode~\acute{c}\else \'{c}\fi{}}, \citenamefont {Kaiser},\ and\
  \citenamefont {Gu{\'e}rin}}]{araujo2016}%
  \BibitemOpen
  \bibfield  {author} {\bibinfo {author} {\bibfnamefont {M.~O.}\ \bibnamefont
  {Ara\'ujo}}, \bibinfo {author} {\bibfnamefont {I.}~\bibnamefont {Kre\ifmmode
  \check{s}\else \v{s}\fi{}i\ifmmode~\acute{c}\else \'{c}\fi{}}}, \bibinfo
  {author} {\bibfnamefont {R.}~\bibnamefont {Kaiser}}, \ and\ \bibinfo {author}
  {\bibfnamefont {W.}~\bibnamefont {Gu{\'e}rin}},\ }\href {\doibase
  10.1103/PhysRevLett.117.073002} {\bibfield  {journal} {\bibinfo  {journal}
  {Phys. Rev. Lett.}\ }\textbf {\bibinfo {volume} {117}},\ \bibinfo {pages}
  {073002} (\bibinfo {year} {2016})}\BibitemShut {NoStop}%
\bibitem [{\citenamefont {Roof}\ \emph {et~al.}(2016)\citenamefont {Roof},
  \citenamefont {Kemp}, \citenamefont {Havey},\ and\ \citenamefont
  {Sokolov}}]{roof2016}%
  \BibitemOpen
  \bibfield  {author} {\bibinfo {author} {\bibfnamefont {S.~J.}\ \bibnamefont
  {Roof}}, \bibinfo {author} {\bibfnamefont {K.~J.}\ \bibnamefont {Kemp}},
  \bibinfo {author} {\bibfnamefont {M.~D.}\ \bibnamefont {Havey}}, \ and\
  \bibinfo {author} {\bibfnamefont {I.~M.}\ \bibnamefont {Sokolov}},\ }\href
  {\doibase 10.1103/PhysRevLett.117.073003} {\bibfield  {journal} {\bibinfo
  {journal} {Phys. Rev. Lett.}\ }\textbf {\bibinfo {volume} {117}},\ \bibinfo
  {pages} {073003} (\bibinfo {year} {2016})}\BibitemShut {NoStop}%
\bibitem [{\citenamefont {Ara\'ujo}\ \emph {et~al.}(2017)\citenamefont
  {Ara\'ujo}, \citenamefont {Gu{\'e}rin},\ and\ \citenamefont
  {Kaiser}}]{araujo2017}%
  \BibitemOpen
  \bibfield  {author} {\bibinfo {author} {\bibfnamefont {M.~O.}\ \bibnamefont
  {Ara\'ujo}}, \bibinfo {author} {\bibfnamefont {W.}~\bibnamefont
  {Gu{\'e}rin}}, \ and\ \bibinfo {author} {\bibfnamefont {R.}~\bibnamefont
  {Kaiser}},\ }\href {\doibase 10.1080/09500340.2017.1380856} {\bibfield
  {journal} {\bibinfo  {journal} {Journal of Modern Optics}\ }\textbf {\bibinfo
  {volume} {65}},\ \bibinfo {pages} {1345} (\bibinfo {year}
  {2017})}\BibitemShut {NoStop}%
\bibitem [{\citenamefont {Sokolov}\ \emph {et~al.}(2009)\citenamefont
  {Sokolov}, \citenamefont {Kupriyanova}, \citenamefont {Kupriyanov},\ and\
  \citenamefont {Havey}}]{sokolov2009}%
  \BibitemOpen
  \bibfield  {author} {\bibinfo {author} {\bibfnamefont {I.~M.}\ \bibnamefont
  {Sokolov}}, \bibinfo {author} {\bibfnamefont {M.~D.}\ \bibnamefont
  {Kupriyanova}}, \bibinfo {author} {\bibfnamefont {D.~V.}\ \bibnamefont
  {Kupriyanov}}, \ and\ \bibinfo {author} {\bibfnamefont {M.~D.}\ \bibnamefont
  {Havey}},\ }\href {\doibase 10.1103/PhysRevA.79.053405} {\bibfield  {journal}
  {\bibinfo  {journal} {Phys. Rev. A}\ }\textbf {\bibinfo {volume} {79}},\
  \bibinfo {pages} {053405} (\bibinfo {year} {2009})}\BibitemShut {NoStop}%
\bibitem [{\citenamefont {Sokolov}\ \emph {et~al.}(2011)\citenamefont
  {Sokolov}, \citenamefont {Kupriyanov},\ and\ \citenamefont
  {Havey}}]{sokolov2011}%
  \BibitemOpen
  \bibfield  {author} {\bibinfo {author} {\bibfnamefont {I.~M.}\ \bibnamefont
  {Sokolov}}, \bibinfo {author} {\bibfnamefont {D.~V.}\ \bibnamefont
  {Kupriyanov}}, \ and\ \bibinfo {author} {\bibfnamefont {M.~D.}\ \bibnamefont
  {Havey}},\ }\href@noop {} {\bibfield  {journal} {\bibinfo  {journal} {J. Exp.
  Theor. Phys.}\ }\textbf {\bibinfo {volume} {112}},\ \bibinfo {pages} {246}
  (\bibinfo {year} {2011})}\BibitemShut {NoStop}%
\bibitem [{\citenamefont {Javanainen}\ \emph {et~al.}(2014)\citenamefont
  {Javanainen}, \citenamefont {Ruostekoski}, \citenamefont {Li},\ and\
  \citenamefont {Yoo}}]{javanainen2014}%
  \BibitemOpen
  \bibfield  {author} {\bibinfo {author} {\bibfnamefont {J.}~\bibnamefont
  {Javanainen}}, \bibinfo {author} {\bibfnamefont {J.}~\bibnamefont
  {Ruostekoski}}, \bibinfo {author} {\bibfnamefont {Y.}~\bibnamefont {Li}}, \
  and\ \bibinfo {author} {\bibfnamefont {S.-M.}\ \bibnamefont {Yoo}},\ }\href
  {\doibase 10.1103/PhysRevLett.112.113603} {\bibfield  {journal} {\bibinfo
  {journal} {Phys. Rev. Lett.}\ }\textbf {\bibinfo {volume} {112}},\ \bibinfo
  {pages} {113603} (\bibinfo {year} {2014})}\BibitemShut {NoStop}%
\bibitem [{\citenamefont {Javanainen}\ and\ \citenamefont
  {Ruostekoski}(2016)}]{javanainen2016}%
  \BibitemOpen
  \bibfield  {author} {\bibinfo {author} {\bibfnamefont {J.}~\bibnamefont
  {Javanainen}}\ and\ \bibinfo {author} {\bibfnamefont {J.}~\bibnamefont
  {Ruostekoski}},\ }\href {\doibase 10.1364/OE.24.000993} {\bibfield  {journal}
  {\bibinfo  {journal} {Opt. Express}\ }\textbf {\bibinfo {volume} {24}},\
  \bibinfo {pages} {993} (\bibinfo {year} {2016})}\BibitemShut {NoStop}%
\bibitem [{\citenamefont {Cherroret}\ \emph {et~al.}(2016)\citenamefont
  {Cherroret}, \citenamefont {Delande},\ and\ \citenamefont {van
  Tiggelen}}]{cherroret2016}%
  \BibitemOpen
  \bibfield  {author} {\bibinfo {author} {\bibfnamefont {N.}~\bibnamefont
  {Cherroret}}, \bibinfo {author} {\bibfnamefont {D.}~\bibnamefont {Delande}},
  \ and\ \bibinfo {author} {\bibfnamefont {B.~A.}\ \bibnamefont {van
  Tiggelen}},\ }\href {\doibase 10.1103/PhysRevA.94.012702} {\bibfield
  {journal} {\bibinfo  {journal} {Phys. Rev. A}\ }\textbf {\bibinfo {volume}
  {94}},\ \bibinfo {pages} {012702} (\bibinfo {year} {2016})}\BibitemShut
  {NoStop}%
\bibitem [{\citenamefont {Friedberg}\ \emph {et~al.}(1973)\citenamefont
  {Friedberg}, \citenamefont {Hartmann},\ and\ \citenamefont
  {Manassah}}]{friedberg1973}%
  \BibitemOpen
  \bibfield  {author} {\bibinfo {author} {\bibfnamefont {R.}~\bibnamefont
  {Friedberg}}, \bibinfo {author} {\bibfnamefont {S.}~\bibnamefont {Hartmann}},
  \ and\ \bibinfo {author} {\bibfnamefont {J.}~\bibnamefont {Manassah}},\
  }\href {\doibase https://doi.org/10.1016/0370-1573(73)90001-X} {\bibfield
  {journal} {\bibinfo  {journal} {Phys. Rep.}\ }\textbf {\bibinfo {volume}
  {7}},\ \bibinfo {pages} {101 } (\bibinfo {year} {1973})}\BibitemShut
  {NoStop}%
\bibitem [{\citenamefont {Keaveney}\ \emph {et~al.}(2012)\citenamefont
  {Keaveney}, \citenamefont {Sargsyan}, \citenamefont {Krohn}, \citenamefont
  {Hughes}, \citenamefont {Sarkisyan},\ and\ \citenamefont
  {Adams}}]{keaveney2012}%
  \BibitemOpen
  \bibfield  {author} {\bibinfo {author} {\bibfnamefont {J.}~\bibnamefont
  {Keaveney}}, \bibinfo {author} {\bibfnamefont {A.}~\bibnamefont {Sargsyan}},
  \bibinfo {author} {\bibfnamefont {U.}~\bibnamefont {Krohn}}, \bibinfo
  {author} {\bibfnamefont {I.~G.}\ \bibnamefont {Hughes}}, \bibinfo {author}
  {\bibfnamefont {D.}~\bibnamefont {Sarkisyan}}, \ and\ \bibinfo {author}
  {\bibfnamefont {C.~S.}\ \bibnamefont {Adams}},\ }\href {\doibase
  10.1103/PhysRevLett.108.173601} {\bibfield  {journal} {\bibinfo  {journal}
  {Phys. Rev. Lett.}\ }\textbf {\bibinfo {volume} {108}},\ \bibinfo {pages}
  {173601} (\bibinfo {year} {2012})}\BibitemShut {NoStop}%
\bibitem [{\citenamefont {Jennewein}\ \emph {et~al.}(2016)\citenamefont
  {Jennewein}, \citenamefont {Besbes}, \citenamefont {Schilder}, \citenamefont
  {Jenkins}, \citenamefont {Sauvan}, \citenamefont {Ruostekoski}, \citenamefont
  {Greffet}, \citenamefont {Sortais},\ and\ \citenamefont
  {Browaeys}}]{jennewein2016}%
  \BibitemOpen
  \bibfield  {author} {\bibinfo {author} {\bibfnamefont {S.}~\bibnamefont
  {Jennewein}}, \bibinfo {author} {\bibfnamefont {M.}~\bibnamefont {Besbes}},
  \bibinfo {author} {\bibfnamefont {N.~J.}\ \bibnamefont {Schilder}}, \bibinfo
  {author} {\bibfnamefont {S.~D.}\ \bibnamefont {Jenkins}}, \bibinfo {author}
  {\bibfnamefont {C.}~\bibnamefont {Sauvan}}, \bibinfo {author} {\bibfnamefont
  {J.}~\bibnamefont {Ruostekoski}}, \bibinfo {author} {\bibfnamefont {J.-J.}\
  \bibnamefont {Greffet}}, \bibinfo {author} {\bibfnamefont {Y.~R.~P.}\
  \bibnamefont {Sortais}}, \ and\ \bibinfo {author} {\bibfnamefont
  {A.}~\bibnamefont {Browaeys}},\ }\href {\doibase
  10.1103/PhysRevLett.116.233601} {\bibfield  {journal} {\bibinfo  {journal}
  {Phys. Rev. Lett.}\ }\textbf {\bibinfo {volume} {116}},\ \bibinfo {pages}
  {233601} (\bibinfo {year} {2016})}\BibitemShut {NoStop}%
\bibitem [{\citenamefont {Bromley}\ \emph {et~al.}(2016)\citenamefont
  {Bromley}, \citenamefont {Zhu}, \citenamefont {Bishof}, \citenamefont
  {Zhang}, \citenamefont {Bothwell}, \citenamefont {Schachenmayer},
  \citenamefont {Nicholson}, \citenamefont {Kaiser}, \citenamefont {Yelin},
  \citenamefont {Lukin}, \citenamefont {Rey},\ and\ \citenamefont
  {Ye}}]{bromley2016}%
  \BibitemOpen
  \bibfield  {author} {\bibinfo {author} {\bibfnamefont {S.~L.}\ \bibnamefont
  {Bromley}}, \bibinfo {author} {\bibfnamefont {B.}~\bibnamefont {Zhu}},
  \bibinfo {author} {\bibfnamefont {M.}~\bibnamefont {Bishof}}, \bibinfo
  {author} {\bibfnamefont {X.}~\bibnamefont {Zhang}}, \bibinfo {author}
  {\bibfnamefont {T.}~\bibnamefont {Bothwell}}, \bibinfo {author}
  {\bibfnamefont {J.}~\bibnamefont {Schachenmayer}}, \bibinfo {author}
  {\bibfnamefont {T.~L.}\ \bibnamefont {Nicholson}}, \bibinfo {author}
  {\bibfnamefont {R.}~\bibnamefont {Kaiser}}, \bibinfo {author} {\bibfnamefont
  {S.~F.}\ \bibnamefont {Yelin}}, \bibinfo {author} {\bibfnamefont {M.~D.}\
  \bibnamefont {Lukin}}, \bibinfo {author} {\bibfnamefont {A.~M.}\ \bibnamefont
  {Rey}}, \ and\ \bibinfo {author} {\bibfnamefont {J.}~\bibnamefont {Ye}},\
  }\href@noop {} {\bibfield  {journal} {\bibinfo  {journal} {Nat. Commun.}\
  }\textbf {\bibinfo {volume} {7}},\ \bibinfo {pages} {11039} (\bibinfo {year}
  {2016})}\BibitemShut {NoStop}%
\bibitem [{\citenamefont {Jennewein}\ \emph {et~al.}(2018)\citenamefont
  {Jennewein}, \citenamefont {Brossard}, \citenamefont {Sortais}, \citenamefont
  {Browaeys}, \citenamefont {Cheinet}, \citenamefont {Robert},\ and\
  \citenamefont {Pillet}}]{jennewein2018}%
  \BibitemOpen
  \bibfield  {author} {\bibinfo {author} {\bibfnamefont {S.}~\bibnamefont
  {Jennewein}}, \bibinfo {author} {\bibfnamefont {L.}~\bibnamefont {Brossard}},
  \bibinfo {author} {\bibfnamefont {Y.~R.~P.}\ \bibnamefont {Sortais}},
  \bibinfo {author} {\bibfnamefont {A.}~\bibnamefont {Browaeys}}, \bibinfo
  {author} {\bibfnamefont {P.}~\bibnamefont {Cheinet}}, \bibinfo {author}
  {\bibfnamefont {J.}~\bibnamefont {Robert}}, \ and\ \bibinfo {author}
  {\bibfnamefont {P.}~\bibnamefont {Pillet}},\ }\href {\doibase
  10.1103/PhysRevA.97.053816} {\bibfield  {journal} {\bibinfo  {journal} {Phys.
  Rev. A}\ }\textbf {\bibinfo {volume} {97}},\ \bibinfo {pages} {053816}
  (\bibinfo {year} {2018})}\BibitemShut {NoStop}%
\bibitem [{\citenamefont {Jenkins}\ \emph {et~al.}(2016)\citenamefont
  {Jenkins}, \citenamefont {Ruostekoski}, \citenamefont {Javanainen},
  \citenamefont {Bourgain}, \citenamefont {Jennewein}, \citenamefont
  {Sortais},\ and\ \citenamefont {Browaeys}}]{jenkins2016}%
  \BibitemOpen
  \bibfield  {author} {\bibinfo {author} {\bibfnamefont {S.~D.}\ \bibnamefont
  {Jenkins}}, \bibinfo {author} {\bibfnamefont {J.}~\bibnamefont
  {Ruostekoski}}, \bibinfo {author} {\bibfnamefont {J.}~\bibnamefont
  {Javanainen}}, \bibinfo {author} {\bibfnamefont {R.}~\bibnamefont
  {Bourgain}}, \bibinfo {author} {\bibfnamefont {S.}~\bibnamefont {Jennewein}},
  \bibinfo {author} {\bibfnamefont {Y.~R.~P.}\ \bibnamefont {Sortais}}, \ and\
  \bibinfo {author} {\bibfnamefont {A.}~\bibnamefont {Browaeys}},\ }\href
  {\doibase 10.1103/PhysRevLett.116.183601} {\bibfield  {journal} {\bibinfo
  {journal} {Phys. Rev. Lett.}\ }\textbf {\bibinfo {volume} {116}},\ \bibinfo
  {pages} {183601} (\bibinfo {year} {2016})}\BibitemShut {NoStop}%
\bibitem [{\citenamefont {Zhu}\ \emph {et~al.}(2016)\citenamefont {Zhu},
  \citenamefont {Cooper}, \citenamefont {Ye},\ and\ \citenamefont
  {Rey}}]{zhu2016}%
  \BibitemOpen
  \bibfield  {author} {\bibinfo {author} {\bibfnamefont {B.}~\bibnamefont
  {Zhu}}, \bibinfo {author} {\bibfnamefont {J.}~\bibnamefont {Cooper}},
  \bibinfo {author} {\bibfnamefont {J.}~\bibnamefont {Ye}}, \ and\ \bibinfo
  {author} {\bibfnamefont {A.~M.}\ \bibnamefont {Rey}},\ }\href@noop {}
  {\bibfield  {journal} {\bibinfo  {journal} {Phys. Rev. A}\ }\textbf {\bibinfo
  {volume} {94}},\ \bibinfo {pages} {023612} (\bibinfo {year}
  {2016})}\BibitemShut {NoStop}%
\bibitem [{\citenamefont {Javanainen}\ \emph {et~al.}(1999)\citenamefont
  {Javanainen}, \citenamefont {Ruostekoski}, \citenamefont {Vestergaard},\ and\
  \citenamefont {Francis}}]{javanainen1999}%
  \BibitemOpen
  \bibfield  {author} {\bibinfo {author} {\bibfnamefont {J.}~\bibnamefont
  {Javanainen}}, \bibinfo {author} {\bibfnamefont {J.}~\bibnamefont
  {Ruostekoski}}, \bibinfo {author} {\bibfnamefont {B.}~\bibnamefont
  {Vestergaard}}, \ and\ \bibinfo {author} {\bibfnamefont {M.~R.}\ \bibnamefont
  {Francis}},\ }\href {\doibase 10.1103/PhysRevA.59.649} {\bibfield  {journal}
  {\bibinfo  {journal} {Phys. Rev. A}\ }\textbf {\bibinfo {volume} {59}},\
  \bibinfo {pages} {649} (\bibinfo {year} {1999})}\BibitemShut {NoStop}%
\bibitem [{\citenamefont {Corman}\ \emph {et~al.}(2017)\citenamefont {Corman},
  \citenamefont {Ville}, \citenamefont {Saint-Jalm}, \citenamefont
  {Aidelsburger}, \citenamefont {Bienaim\'e}, \citenamefont {Nascimb\`ene},
  \citenamefont {Dalibard},\ and\ \citenamefont {Beugnon}}]{corman2017}%
  \BibitemOpen
  \bibfield  {author} {\bibinfo {author} {\bibfnamefont {L.}~\bibnamefont
  {Corman}}, \bibinfo {author} {\bibfnamefont {J.~L.}\ \bibnamefont {Ville}},
  \bibinfo {author} {\bibfnamefont {R.}~\bibnamefont {Saint-Jalm}}, \bibinfo
  {author} {\bibfnamefont {M.}~\bibnamefont {Aidelsburger}}, \bibinfo {author}
  {\bibfnamefont {T.}~\bibnamefont {Bienaim\'e}}, \bibinfo {author}
  {\bibfnamefont {S.}~\bibnamefont {Nascimb\`ene}}, \bibinfo {author}
  {\bibfnamefont {J.}~\bibnamefont {Dalibard}}, \ and\ \bibinfo {author}
  {\bibfnamefont {J.}~\bibnamefont {Beugnon}},\ }\href {\doibase
  10.1103/PhysRevA.96.053629} {\bibfield  {journal} {\bibinfo  {journal} {Phys.
  Rev. A}\ }\textbf {\bibinfo {volume} {96}},\ \bibinfo {pages} {053629}
  (\bibinfo {year} {2017})}\BibitemShut {NoStop}%
\bibitem [{\citenamefont {van Tiggelen}\ and\ \citenamefont
  {Lagendijk}(1994)}]{vantiggelen1994}%
  \BibitemOpen
  \bibfield  {author} {\bibinfo {author} {\bibfnamefont {B.~A.}\ \bibnamefont
  {van Tiggelen}}\ and\ \bibinfo {author} {\bibfnamefont {A.}~\bibnamefont
  {Lagendijk}},\ }\href {\doibase 10.1103/PhysRevB.50.16729} {\bibfield
  {journal} {\bibinfo  {journal} {Phys. Rev. B}\ }\textbf {\bibinfo {volume}
  {50}},\ \bibinfo {pages} {16729} (\bibinfo {year} {1994})}\BibitemShut
  {NoStop}%
\bibitem [{\citenamefont {Wiersma}\ \emph {et~al.}(1995)\citenamefont
  {Wiersma}, \citenamefont {van Albada}, \citenamefont {van Tiggelen},\ and\
  \citenamefont {Lagendijk}}]{wiersma1995}%
  \BibitemOpen
  \bibfield  {author} {\bibinfo {author} {\bibfnamefont {D.~S.}\ \bibnamefont
  {Wiersma}}, \bibinfo {author} {\bibfnamefont {M.~P.}\ \bibnamefont {van
  Albada}}, \bibinfo {author} {\bibfnamefont {B.~A.}\ \bibnamefont {van
  Tiggelen}}, \ and\ \bibinfo {author} {\bibfnamefont {A.}~\bibnamefont
  {Lagendijk}},\ }\href@noop {} {\bibfield  {journal} {\bibinfo  {journal}
  {Phys. Rev. Lett.}\ }\textbf {\bibinfo {volume} {74}},\ \bibinfo {pages}
  {4193} (\bibinfo {year} {1995})}\BibitemShut {NoStop}%
\bibitem [{\citenamefont {Pierrat}\ and\ \citenamefont
  {Carminati}(2010)}]{pierrat2010}%
  \BibitemOpen
  \bibfield  {author} {\bibinfo {author} {\bibfnamefont {R.}~\bibnamefont
  {Pierrat}}\ and\ \bibinfo {author} {\bibfnamefont {R.}~\bibnamefont
  {Carminati}},\ }\href@noop {} {\bibfield  {journal} {\bibinfo  {journal}
  {Phys. Rev. A}\ }\textbf {\bibinfo {volume} {81}},\ \bibinfo {pages} {063802}
  (\bibinfo {year} {2010})}\BibitemShut {NoStop}%
\bibitem [{\citenamefont {Aubry}\ \emph {et~al.}(2014)\citenamefont {Aubry},
  \citenamefont {Cobus}, \citenamefont {Skipetrov}, \citenamefont {van
  Tiggelen}, \citenamefont {Derode},\ and\ \citenamefont {Page}}]{aubry2014}%
  \BibitemOpen
  \bibfield  {author} {\bibinfo {author} {\bibfnamefont {A.}~\bibnamefont
  {Aubry}}, \bibinfo {author} {\bibfnamefont {L.~A.}\ \bibnamefont {Cobus}},
  \bibinfo {author} {\bibfnamefont {S.~E.}\ \bibnamefont {Skipetrov}}, \bibinfo
  {author} {\bibfnamefont {B.~A.}\ \bibnamefont {van Tiggelen}}, \bibinfo
  {author} {\bibfnamefont {A.}~\bibnamefont {Derode}}, \ and\ \bibinfo {author}
  {\bibfnamefont {J.~H.}\ \bibnamefont {Page}},\ }\href@noop {} {\bibfield
  {journal} {\bibinfo  {journal} {Phys. Rev. Lett.}\ }\textbf {\bibinfo
  {volume} {112}},\ \bibinfo {pages} {043903} (\bibinfo {year}
  {2014})}\BibitemShut {NoStop}%
\bibitem [{\citenamefont {Frisch}(1968)}]{frisch1968}%
  \BibitemOpen
  \bibfield  {author} {\bibinfo {author} {\bibfnamefont {U.}~\bibnamefont
  {Frisch}},\ }in\ \href@noop {} {\emph {\bibinfo {booktitle} {Probabilistic
  Methods in Applied Mathematics, Volume 1}}},\ \bibinfo {editor} {edited by\
  \bibinfo {editor} {\bibfnamefont {A.~T.}\ \bibnamefont {Baharucha-Reid}}}\
  (\bibinfo  {publisher} {Academic Press, Inc., New York},\ \bibinfo {year}
  {1968})\BibitemShut {NoStop}%
\bibitem [{\citenamefont {Morice}\ \emph {et~al.}(1995)\citenamefont {Morice},
  \citenamefont {Castin},\ and\ \citenamefont {Dalibard}}]{morice1995}%
  \BibitemOpen
  \bibfield  {author} {\bibinfo {author} {\bibfnamefont {O.}~\bibnamefont
  {Morice}}, \bibinfo {author} {\bibfnamefont {Y.}~\bibnamefont {Castin}}, \
  and\ \bibinfo {author} {\bibfnamefont {J.}~\bibnamefont {Dalibard}},\
  }\href@noop {} {\bibfield  {journal} {\bibinfo  {journal} {Phys. Rev. A}\
  }\textbf {\bibinfo {volume} {51}},\ \bibinfo {pages} {3896} (\bibinfo {year}
  {1995})}\BibitemShut {NoStop}%
\bibitem [{\citenamefont {Ruostekoski}\ and\ \citenamefont
  {Javanainen}(1997)}]{ruostekoski1997}%
  \BibitemOpen
  \bibfield  {author} {\bibinfo {author} {\bibfnamefont {J.}~\bibnamefont
  {Ruostekoski}}\ and\ \bibinfo {author} {\bibfnamefont {J.}~\bibnamefont
  {Javanainen}},\ }\href@noop {} {\bibfield  {journal} {\bibinfo  {journal}
  {Phys. Rev. A}\ }\textbf {\bibinfo {volume} {55}},\ \bibinfo {pages} {513}
  (\bibinfo {year} {1997})}\BibitemShut {NoStop}%
\bibitem [{\citenamefont {Ruostekoski}\ and\ \citenamefont
  {Javanainen}(1999)}]{ruostekoski1999}%
  \BibitemOpen
  \bibfield  {author} {\bibinfo {author} {\bibfnamefont {J.}~\bibnamefont
  {Ruostekoski}}\ and\ \bibinfo {author} {\bibfnamefont {J.}~\bibnamefont
  {Javanainen}},\ }\href {\doibase 10.1103/PhysRevLett.82.4741} {\bibfield
  {journal} {\bibinfo  {journal} {Phys. Rev. Lett.}\ }\textbf {\bibinfo
  {volume} {82}},\ \bibinfo {pages} {4741} (\bibinfo {year}
  {1999})}\BibitemShut {NoStop}%
\bibitem [{\citenamefont {Pierrat}\ \emph {et~al.}(2009)\citenamefont
  {Pierrat}, \citenamefont {Gr\'emaud},\ and\ \citenamefont
  {Delande}}]{pierrat2009}%
  \BibitemOpen
  \bibfield  {author} {\bibinfo {author} {\bibfnamefont {R.}~\bibnamefont
  {Pierrat}}, \bibinfo {author} {\bibfnamefont {B.}~\bibnamefont {Gr\'emaud}},
  \ and\ \bibinfo {author} {\bibfnamefont {D.}~\bibnamefont {Delande}},\ }\href
  {\doibase 10.1103/PhysRevA.80.013831} {\bibfield  {journal} {\bibinfo
  {journal} {Phys. Rev. A}\ }\textbf {\bibinfo {volume} {80}},\ \bibinfo
  {pages} {013831} (\bibinfo {year} {2009})}\BibitemShut {NoStop}%
\bibitem [{\citenamefont {Percus}\ and\ \citenamefont
  {Yevick}(1958)}]{percus1958}%
  \BibitemOpen
  \bibfield  {author} {\bibinfo {author} {\bibfnamefont {J.~K.}\ \bibnamefont
  {Percus}}\ and\ \bibinfo {author} {\bibfnamefont {G.~J.}\ \bibnamefont
  {Yevick}},\ }\href@noop {} {\bibfield  {journal} {\bibinfo  {journal} {Phys.
  Rev.}\ }\textbf {\bibinfo {volume} {110}},\ \bibinfo {pages} {1} (\bibinfo
  {year} {1958})}\BibitemShut {NoStop}%
\bibitem [{\citenamefont {Wertheim}(1963)}]{wertheim1963}%
  \BibitemOpen
  \bibfield  {author} {\bibinfo {author} {\bibfnamefont {M.~S.}\ \bibnamefont
  {Wertheim}},\ }\href@noop {} {\bibfield  {journal} {\bibinfo  {journal}
  {Phys. Rev. Lett.}\ }\textbf {\bibinfo {volume} {10}},\ \bibinfo {pages}
  {321} (\bibinfo {year} {1963})}\BibitemShut {NoStop}%
\bibitem [{\citenamefont {Bons}\ \emph {et~al.}(2016)\citenamefont {Bons},
  \citenamefont {de~Haas}, \citenamefont {de~Jong}, \citenamefont {Groot},\
  and\ \citenamefont {van~der Straten}}]{bons2016}%
  \BibitemOpen
  \bibfield  {author} {\bibinfo {author} {\bibfnamefont {P.~C.}\ \bibnamefont
  {Bons}}, \bibinfo {author} {\bibfnamefont {R.}~\bibnamefont {de~Haas}},
  \bibinfo {author} {\bibfnamefont {D.}~\bibnamefont {de~Jong}}, \bibinfo
  {author} {\bibfnamefont {A.}~\bibnamefont {Groot}}, \ and\ \bibinfo {author}
  {\bibfnamefont {P.}~\bibnamefont {van~der Straten}},\ }\href {\doibase
  10.1103/PhysRevLett.116.173602} {\bibfield  {journal} {\bibinfo  {journal}
  {Phys. Rev. Lett.}\ }\textbf {\bibinfo {volume} {116}},\ \bibinfo {pages}
  {173602} (\bibinfo {year} {2016})}\BibitemShut {NoStop}%
\bibitem [{\citenamefont {Allen}\ and\ \citenamefont
  {Eberly}(1974)}]{allen1974}%
  \BibitemOpen
  \bibfield  {author} {\bibinfo {author} {\bibfnamefont {L.}~\bibnamefont
  {Allen}}\ and\ \bibinfo {author} {\bibfnamefont {J.~H.}\ \bibnamefont
  {Eberly}},\ }\href@noop {} {\emph {\bibinfo {title} {Optical resonance and
  two-level atoms}}}\ (\bibinfo  {publisher} {Dover Publication, New York},\
  \bibinfo {year} {1974})\BibitemShut {NoStop}%
\bibitem [{\citenamefont {Akkermans}\ and\ \citenamefont
  {Montambaux}(2007)}]{akkermans2007}%
  \BibitemOpen
  \bibfield  {author} {\bibinfo {author} {\bibfnamefont {E.}~\bibnamefont
  {Akkermans}}\ and\ \bibinfo {author} {\bibfnamefont {G.}~\bibnamefont
  {Montambaux}},\ }\href@noop {} {\emph {\bibinfo {title} {Mesoscopic Physics
  of Electrons and Photons}}}\ (\bibinfo  {publisher} {Cambridge University
  Press, Cambridge},\ \bibinfo {year} {2007})\BibitemShut {NoStop}%
\bibitem [{\citenamefont {Dyson}(1949{\natexlab{a}})}]{dyson1949a}%
  \BibitemOpen
  \bibfield  {author} {\bibinfo {author} {\bibfnamefont {F.~J.}\ \bibnamefont
  {Dyson}},\ }\href@noop {} {\bibfield  {journal} {\bibinfo  {journal} {Phys.
  Rev.}\ }\textbf {\bibinfo {volume} {75}},\ \bibinfo {pages} {486} (\bibinfo
  {year} {1949}{\natexlab{a}})}\BibitemShut {NoStop}%
\bibitem [{\citenamefont {Dyson}(1949{\natexlab{b}})}]{dyson1949b}%
  \BibitemOpen
  \bibfield  {author} {\bibinfo {author} {\bibfnamefont {F.~J.}\ \bibnamefont
  {Dyson}},\ }\href@noop {} {\bibfield  {journal} {\bibinfo  {journal} {Phys.
  Rev.}\ }\textbf {\bibinfo {volume} {75}},\ \bibinfo {pages} {1736} (\bibinfo
  {year} {1949}{\natexlab{b}})}\BibitemShut {NoStop}%
\bibitem [{\citenamefont {Lagendijk}\ and\ \citenamefont {van
  Tiggelen}(1996)}]{lagendijk1996}%
  \BibitemOpen
  \bibfield  {author} {\bibinfo {author} {\bibfnamefont {A.}~\bibnamefont
  {Lagendijk}}\ and\ \bibinfo {author} {\bibfnamefont {B.~A.}\ \bibnamefont
  {van Tiggelen}},\ }\href@noop {} {\bibfield  {journal} {\bibinfo  {journal}
  {Phys. Rep.}\ }\textbf {\bibinfo {volume} {270}},\ \bibinfo {pages} {143}
  (\bibinfo {year} {1996})}\BibitemShut {NoStop}%
\bibitem [{\citenamefont {Rytov}\ \emph {et~al.}(1989)\citenamefont {Rytov},
  \citenamefont {Kravtsov},\ and\ \citenamefont {Tatarskii}}]{rytov1989}%
  \BibitemOpen
  \bibfield  {author} {\bibinfo {author} {\bibfnamefont {S.~M.}\ \bibnamefont
  {Rytov}}, \bibinfo {author} {\bibfnamefont {Y.~A.}\ \bibnamefont {Kravtsov}},
  \ and\ \bibinfo {author} {\bibfnamefont {V.~I.}\ \bibnamefont {Tatarskii}},\
  }\href@noop {} {\emph {\bibinfo {title} {Principles of Statistical
  Radiophysics IV}}}\ (\bibinfo  {publisher} {Springer-Verlag, Berlin},\
  \bibinfo {year} {1989})\BibitemShut {NoStop}%
\bibitem [{\citenamefont {Tai}(1993)}]{tai1993}%
  \BibitemOpen
  \bibfield  {author} {\bibinfo {author} {\bibfnamefont {C.-T.}\ \bibnamefont
  {Tai}},\ }\href@noop {} {\emph {\bibinfo {title} {Dyadic Green Functions in
  Electromagnetic Theory}}}\ (\bibinfo  {publisher} {IEEE Press, New York},\
  \bibinfo {year} {1993})\BibitemShut {NoStop}%
\bibitem [{\citenamefont {Mallet}\ \emph {et~al.}(2005)\citenamefont {Mallet},
  \citenamefont {Gu{\'e}rin},\ and\ \citenamefont {Sentenac}}]{mallet2005}%
  \BibitemOpen
  \bibfield  {author} {\bibinfo {author} {\bibfnamefont {P.}~\bibnamefont
  {Mallet}}, \bibinfo {author} {\bibfnamefont {C.~A.}\ \bibnamefont
  {Gu{\'e}rin}}, \ and\ \bibinfo {author} {\bibfnamefont {A.}~\bibnamefont
  {Sentenac}},\ }\href@noop {} {\bibfield  {journal} {\bibinfo  {journal}
  {Phys. Rev. B}\ }\textbf {\bibinfo {volume} {72}},\ \bibinfo {pages} {014205}
  (\bibinfo {year} {2005})}\BibitemShut {NoStop}%
\bibitem [{\citenamefont {Gu{\'e}rin}\ \emph {et~al.}(2006)\citenamefont
  {Gu{\'e}rin}, \citenamefont {Mallet},\ and\ \citenamefont
  {Sentenac}}]{guerin2006}%
  \BibitemOpen
  \bibfield  {author} {\bibinfo {author} {\bibfnamefont {C.-A.}\ \bibnamefont
  {Gu{\'e}rin}}, \bibinfo {author} {\bibfnamefont {P.}~\bibnamefont {Mallet}},
  \ and\ \bibinfo {author} {\bibfnamefont {A.}~\bibnamefont {Sentenac}},\
  }\href@noop {} {\bibfield  {journal} {\bibinfo  {journal} {J. Opt. Soc. Am.
  A}\ }\textbf {\bibinfo {volume} {23}},\ \bibinfo {pages} {349} (\bibinfo
  {year} {2006})}\BibitemShut {NoStop}%
\bibitem [{\citenamefont {Born}\ and\ \citenamefont {Wolf}(1999)}]{born1999}%
  \BibitemOpen
  \bibfield  {author} {\bibinfo {author} {\bibfnamefont {M.}~\bibnamefont
  {Born}}\ and\ \bibinfo {author} {\bibfnamefont {E.}~\bibnamefont {Wolf}},\
  }\href@noop {} {\emph {\bibinfo {title} {Principles of Optics}}}\ (\bibinfo
  {publisher} {Cambridge University Press, Cambridge, UK},\ \bibinfo {year}
  {1999})\BibitemShut {NoStop}%
\bibitem [{\citenamefont {Naraschewski}\ and\ \citenamefont
  {Glauber}(1999)}]{naraschewski1999}%
  \BibitemOpen
  \bibfield  {author} {\bibinfo {author} {\bibfnamefont {M.}~\bibnamefont
  {Naraschewski}}\ and\ \bibinfo {author} {\bibfnamefont {R.~J.}\ \bibnamefont
  {Glauber}},\ }\href {\doibase 10.1103/PhysRevA.59.4595} {\bibfield  {journal}
  {\bibinfo  {journal} {Phys. Rev. A}\ }\textbf {\bibinfo {volume} {59}},\
  \bibinfo {pages} {4595} (\bibinfo {year} {1999})}\BibitemShut {NoStop}%
\end{thebibliography}
\end{document}